%% file: apssamp.tex
\documentclass[%
 reprint,
superscriptaddress,
 amsmath,amssymb,
 aps,
floatfix,
]{revtex4-2}

\usepackage{graphicx, subcaption}
\graphicspath{ {./images3/} }
\usepackage{dcolumn}
\usepackage{bm}
\usepackage{gensymb}
\usepackage{rotating}
\usepackage{booktabs}
\usepackage{color}

\begin{document}


\title{Pipelined information flow in molecular mechanical circuits leads to increased error and irreversibility}

\author{Ian Seet}
\affiliation{ 
Physical and Theoretical Chemistry Laboratory, Department of Chemistry,University of Oxford, South Parks Road, Oxford OX1 3QZ, United Kingdom
}%

\author{Thomas E. Ouldridge}
\affiliation{ 
Department of Bioengineering,
Imperial College London,
Royal School of Mines,
Exhibition Road,
London,
SW7 2AZ,
United Kingdom
}%

\author{Jonathan P.K. Doye}
\affiliation{ 
Physical and Theoretical Chemistry Laboratory, Department of Chemistry,University of Oxford, South Parks Road, Oxford OX1 3QZ, United Kingdom
}%

\date{\today}

\begin{abstract}
Pipelining is a design technique for logical circuits that allows for higher throughput than circuits in which multiple computations are fed through the system one after the other. It allows for much faster computation than architectures in which inputs must pass through every layer of the circuit before the next computation can begin (phased chaining). We explore the hypothesis that these advantages may be offset by a higher error rate, logical irreversibility and greater thermodynamic costs by simulating pipelined molecular mechanical circuits using an explicit physical model. We observe the emergent logical irreversibility, and see that the simultaneous action of multiple components indeed leads to a higher error rate than in phase-chained circuits. The thermodynamic costs of operating the gates are much larger than in equivalent phase-chained circuits, and these costs do not appear to tend to zero in the limit of slow gate operation. Redesigning the gates to eliminate errors and artificially enforcing logical reversibility reduces the thermodynamic costs and recovers thermodynamically reversible behaviour in the limit of slow gate operation. The breakdown of logical reversibility and accuracy are both associated with a breakdown of the digital behaviour of the device, likely contributing to thermodynamic costs that are large relative to the scale of the information being processed.
\end{abstract}

\maketitle


\section{\label{sec:intro}Introduction}
Much of the historical work on the thermodynamics of computation has focused on individual bits \cite{Lan61, Fred82}, including Landauer's celebrated observation that bit erasure requires a minimum of $k_\mathrm{B}T\ln{2}$ of work input \cite{Lan61}. Extending these ideas to more practically relevant computations is an active field of research; for example, the Landauer work cost can be generalised to more complex logically irreversible computations, in which the input state cannot be unambiguously inferred from the output state \cite{wol19}.

In theory, this generalized Landauer work cost does not necessarily increase the total entropy of the universe 
\cite{fahn96, mar05}, and the work can be recovered in a well-designed system that operates thermodynamically reversibly. In practice, however, performing a complicated logically irreversible operation in a thermodynamically reversible manner would be extremely challenging \cite{Brit21,ould22}. Indeed, a far-ranging recent result is that logically irreversible operations are subject to a ``mismatch cost'' \cite{wol19,wol20}. The mismatch cost sets a non-zero lower bound on the entropy production, and hence the irrecoverable thermodynamic cost, associated with a logically irreversible operation, unless the device is designed specifically to operate only on the narrow range of initial distributions of inputs that are optimal. 


Results such as the mismatch cost are powerful because they are general and do not assume much about the underlying architecture. However, this approach has limitations, in that it is not always clear when these costs will arise, whether they will be a significant factor in the computational costs, and whether other factors or trade-offs will come into play.

A complementary approach is to explore the thermodynamics of particular computational architectures. For example, combinatorial circuits provide a pathway to exploring the thermodynamics of computation in simple settings in which non-trivial computations are performed \cite{wol19,wol20}. Combinatorial circuits are digital logic circuits composed of multiple Boolean logic gates in which all possible outputs of the function are influenced only by the corresponding input state and not any previous state the system has taken.
In previous work, we have outlined an explicit model that allows for the simulation of  molecular mechanical logic gates and combinatorial circuits \cite{seet22}. Since our framework allows for the simulation of well-defined (albeit idealised) physical systems, it allows for a more thorough exploration of how abstract constraints feed into a physical system \cite{Kol20}.

The molecular mechanical logic gates we have outlined are distinguished from existing reversible molecular logic gate designs that exploit the specificity of DNA base-pairing \cite{Gen11, Tao14} by their reliance on mechanical means to transmit information as opposed to relying on bond-breaking and diffusional processes, resulting in a potential large increase in speed. Conversely, these gates differ from existing mechanical logic gate designs \cite{Yas21, Mer93, Wen11, Mer18, Song19} by their far lower mass, such that their kinetic energy remains close to equipartition even when operating at high clock speeds.

In Ref.~\cite{seet22}, we simulated large, logically and thermodynamically reversible combinatorial circuits constructed from the aforementioned logic gates via a protocol that we have termed ``phased chaining'', in which each level of logic gates in a combinatorial circuit completes its operation before the next level is initiated. 
Although this protocol was successful at reducing the error rate in comparison to direct mechanical coupling of logic gates, while also remaining thermodynamically reversible, it requires the use of multiple external controls to drive the system, one for each distinct level of the circuit. In addition, phased chaining also necessitates that every gate in the circuit process its input and return its output before any upstream gates can be released to accept new inputs. As a result, systems employing phased chaining both have a low throughput and are difficult to scale.

A simple method to reduce the complexity and increase the speed of a combinatorial circuit is to drive the system via a single external driver, so that all gates are operating at once, while maintaining the staggering of phases. This technique, known as pipelining \cite{bev04}, allows each gate in the chain to simultaneously process information once per clock cycle, as opposed to the phased-chaining protocol where each gate can only process information once during the time it takes for information to pass from the first level of the circuit to the last, regardless of how many clock cycles are required.

In this paper, we will explore pipelining within our framework of molecular mechanical logic gates. As our framework currently lacks any method of preserving information beyond a single clock cycle, the pipelining implementation used in this paper more closely resembles the wave-pipelining approach \cite{Wong93, Bur98, gray94} than the standard register-based pipelining approach described in circuit design textbooks, which does require persistent information storage (sequential logic) \cite{shen05}. We will, however, refer to the protocol used in this paper as pipelining for the sake of simplicity.

A phase-chained circuit with gates that preserve the inputs at each layer of the circuit until the end of the computation is intrinsically logically reversible. A pipelined system does not possess this intrinsic logical reversibility. In the absence of a mechanism that enables or simulates the storage of information beyond a single clock cycle, a pipelined circuit will, in general, be logically irreversible. We therefore expect that pipelined circuits will be subject to a mismatch cost, and hence in general more thermodynamically costly than phase-chained circuits. We also expect that, since many layers of a pipelined system operate in parallel, rather than only a single layer with all others fixed in a phase-chained system, that logical errors will be more likely. 

In this work we test these hypotheses, showing that logical irreversibility and enhanced errors are indeed a consequence of pipelining. Pipelined circuits with logical irreversibility and high error rate exhibit high entropy production and hence large thermodynamic costs, which can be suppressed by gate redesign and artificially imposing logical irreversibility. We find that the loss of logical reversibility and accuracy are both associated with a breakdown of the digital behaviour of the device, so that the details of the microscopic state of the system become important. We posit that this breakdown of digital behaviour contributes significantly to thermodynamic costs that are large relative to the scale of the information being processed.

In Section \ref{sec:irrevPipe}, we first demonstrate a simple pipelined system consisting of two bits, and show that failing to preserve logical reversibility has a large negative effect on thermodynamic efficiency. In Section \ref{sec:revPipe}, we then show how such a circuit may be rendered logically and thermodynamically reversible by simulating the effects of information storage. Subsequently, we attempt to demonstrate a similar technique on a more complex system consisting of up to four pipelined NAND gates, and observe that the high error rate associated with a larger number of moving parts can cause a large decrease in thermodynamic efficiency. Finally, we show that by reducing the error rate with an improved driver design, we can restore thermodynamic reversibility to the system.




\section{\label{sec:rigormortis}Methods of Simulation and Circuit Components}
A rigid-body molecular dynamics simulator \cite{Dav14} identical in all parameters to the one used in our previous work \cite{seet22} was used to simulate all logic gates and circuits in this paper. The logic gates and other circuit components used to construct the pipelined circuit were also drawn from this work, and a similar mechanism is used to drive the circuits. The specific circuit components relevant to the pipelined circuit are the driver, winged NAND gate and switch. Succinct summaries of the mode of actions of these components are provided here; full details can be found in Ref.\ \cite{seet22}.

\begin{figure*}
\includegraphics[width=\linewidth]{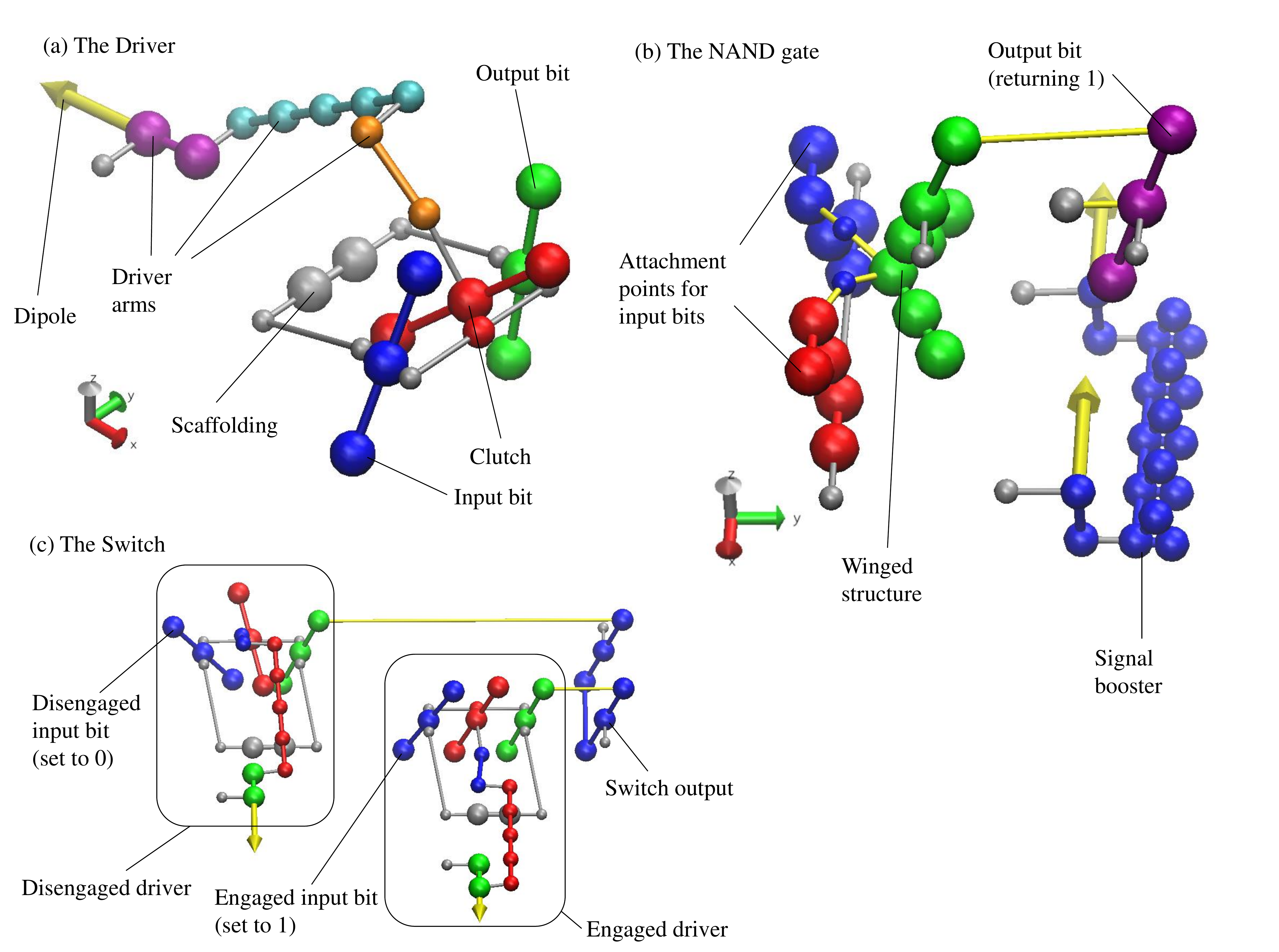}
\caption{The three critical circuit components from which the pipelined circuits are constructed. The differently-sized spheres represent the three different particle types used. Silver-coloured bonds represent bonding interactions with an angular component between rigid bodies, while yellow-coloured bonds represent purely distance-based interactions. The silver spheres represent the scaffold to which the rigid bodies are constrained. The scaffold particles are fixed in space. Particles directly bound to the scaffold are constrained by a harmonic potential centered on their initial position. 
The magnitudes of the harmonic constants for the bonding and angular potentials used are identical to our previous work \cite{seet22}. The yellow arrow indicates an internal dipole that interacts with the clock-driven external dipole via the potential of Eq.\ \ref{eq:dipole}.
A similar colouring scheme is used for the diagrams throughout the paper.}
\label{fig:cclabel}
\end{figure*}

The molecular mechanical circuits we use to demonstrate pipelining are constructed from rigid bodies which are in turn composed of three distinct types of spherical particle, corresponding to masses of 120, 41 and 14 amu. The rigid bodies are connected by bonds (which include both a distance-based and angle-based harmonic potential, coloured silver in Fig.\ref{fig:cclabel}) and tethers (which only include a distance-based potential, coloured yellow in Fig. \ref{fig:cclabel}).

The work required to drive the circuits originates from one or more rotating external dipoles which interact with internal dipoles in the system, and information transduction between gates is achieved through mechanical force transmission only. The circuits are maintained near equipartition (a necessary condition for molecular mechanical logic) by a Langevin thermostat.

\subsection{\label{ssec:extDip}Driving the logic gates}
All systems in this paper are driven by one or more external dipoles that rotate at constant angular velocity until the computation is reversed, at which point the angular velocity is negated. These dipoles couple to internal dipoles, with the dipole-dipole potential taking the form
\begin{eqnarray}
V_{\mathrm{dipole}} = k_{\mathrm{dip}}\mathbf{\hat{d}_{\mathrm{ext}}}\cdot\mathbf{\hat{d}_{\mathrm{int}}},
\label{eq:dipole}
\end{eqnarray}
where $\mathbf{\hat{d}_{\mathrm{ext}}}$ and $\mathbf{\hat{d}_{\mathrm{int}}}$ are the direction vectors of the external and internal system dipoles, respectively. 
Each actively driven circuit component contains internal dipoles that interact with the external dipoles independently. The following expression \cite{Sek97, Sek98} was used to calculate the work done on the gate by the driving force of a single dipole:
\begin{eqnarray}
W_{\mathrm{dipole}}(t, {\Delta}t) = k_{\mathrm{dip}}(\mathbf{\hat{d}_{\mathrm{ext}}}(t + {\Delta}t) - \mathbf{\hat{d}_{\mathrm{ext}}}(t)) \cdot\mathbf{\hat{d}_{\mathrm{int}}},
\label{eq:work}
\end{eqnarray}
where $\mathbf{\hat{d}_{\mathrm{ext}}}(t)$ is the direction vector of the external dipole at time $t$, ${\Delta}t$ is the length of one timestep, and $W_{\mathrm{dipole}}$ the work done by the external dipole on a given internal dipole across the timestep ${\Delta}t$. This expression equates work done to the increase in energy due to the change in the control parameter and has been used to calculate the work done in experimental realisations of the Landauer limit\,\cite{jun14}.

For the purposes of this paper, we explore thermodynamic reversibility by applying a control protocol to the external dipoles for the first half of a simulation, then reversing that protocol for the second half. If the system returns to its initial distribution of states with zero total work input, then the applied protocol is thermodynamically reversible. In principle, thermodynamic reversibility can only be achieved in the limit of quasistatic (infinitely slow) manipulations; we describe a system as behaving in a thermodynamically reversible way if the entropy production tends towards zero linearly with the clock speed (the inverse of the duration of the protocol). We note that, since the simulations are in a strongly damped regime, it was unnecessary to reverse the momenta of all particles prior to starting the reverse protocols as the average kinetic energy of the system does not significantly exceed the kinetic energy predicted by equipartition.

It is important to emphasize that the dipoles used here merely constitute a well-defined and limited way in which an externally applied protocol can couple to the computational degrees of freedom of the system and push the computation forwards. They should not be interpreted any more deeply than that. A single control signal is all that is required for each set of external dipoles that rotate at a fixed relative phase. By contrast, distinct control signals are required for external dipoles that operate independently from each other.

\subsection{\label{ssec:driver}The Driver}
The purpose of the driver is to selectively transduce information between an input bit and an output bit depending on the state of the external dipole. Fig.\ \ref{fig:cclabel}(a) illustrates the driver design as well as the colouring scheme used in all circuits in this paper. 

This selectivity is implemented by altering the position and orientation of the red clutch based on the state of the internal dipole. When the driver is in a state of maximal engagement, the internal dipole of the driver pulls the clutch into a position where it can transfer via steric interactions the state of the blue input bit to the green output bit.

At the position of maximal disengagement, the clutch is forced into a position that decouples the output and input bits from each other. For all structures discussed in this paper, an input bit with a positive slope in the $yz$ plane has a value of `1' and an input bit with a negative slope has a value of `0'; thus, the input bit of the driver in Fig. \ref{fig:cclabel}(a) is currently in the `1' state.
\subsection{\label{ssec:nand}The NAND gate}
Any combinatorial Boolean logic circuit may be constructed from NAND gates alone. The molecular mechanical NAND gate outlined in Fig.\ \ref{fig:cclabel}(b) was designed to function as the basic logic gate for both phase-chained and pipelined circuits. The output bit of the gate is connected to a freely-rotating pair of prong- or wing-like structures, hence the term ``winged'' NAND gate. The tip of the output bit is tethered to the tips of the input bits via the winged structure.

When the input bits are fixed to the state `00', the winged structure is pulled by the input bits into the `1' configuration. When the input bits are set to `01' or `10', the winged structure is simultaneously pulled by the `0'-set bit while being pushed by the `1'-set bit. Due to its ability to pivot away from the input bits, the winged structure is forced into the `1' configuration by these opposing sets of forces. When the input bits are set to `11', both bits push against the winged structure and it is forced into the `0' configuration.

The signal booster is a wedge-shaped block that is driven by the external dipoles. During a state of maximal engagement, it is forced toward the output bit. The steric interaction betwen the signal booster and output bit forces the output bit away from the $z$-axis and thereby increases the deflection of the output bit from that axis, effectively boosting the signal. Moving the signal booster is the equivalent of raising/lowering the barrier between states in more conventional approaches to modelling computation \cite{ould17}.

\subsection{\label{ssec:switch}The Switch}
The switch provides a means of selectively relaying information to an output from two different input bits via a change in the phase of the external dipole. It consists of two drivers that have a phase offset of 180\textdegree\ such that the internal dipoles of the two drivers point in opposite directions and that are tethered to a common switch output. By altering the phase of the external dipole, one of the input bits of the drivers can be selectively expressed at the output (Fig.\ \ref{fig:cclabel}(c)).

By connecting two pairs of switches to the NAND gate, the inputs of the NAND gate can be selectively varied as a function of the phase of the external dipole. This mechanism allows for time-dependent transitions between arbitrary pairs of states.

\begin{figure}
	\includegraphics[width=\linewidth]{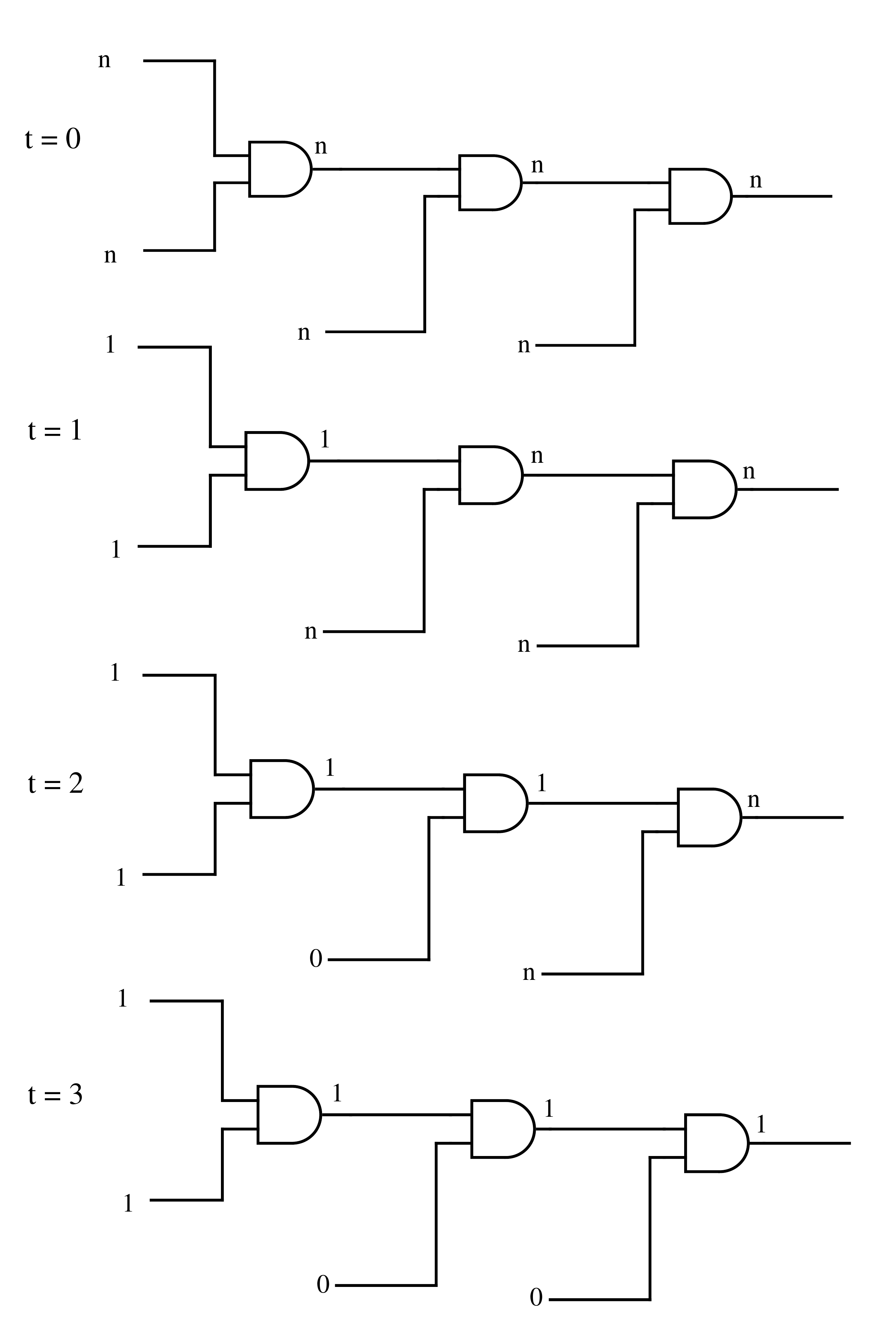}
	\caption{A simplified demonstration of phased chaining in a circuit containing three chained AND gates. Units of time are arbitrary. States labelled `n' are neutral disengaged states set to neither `0' or `1'. The figure illustrates the three steps needed to compute a given output when the initial input is delivered to the system at $t = 1$.}
	\label{fig:phasedChaining}
\end{figure}

\subsection{\label{ssec:phasedChain}Phased Chaining vs. Pipelining}
We have used the concept of phased chaining in combinatorial circuits in order to achieve logical and thermodynamic reversibility while minimising the error rate \cite{seet22}. In phased chaining, each level of logic gates is engaged only after the previous level has become engaged. A logic gate will also never become disengaged until every gate in every level ahead of it has become disengaged. Thus, in the case of the circuit depicted in Fig.\ \ref{fig:phasedChaining}, should the output of the circuit given inputs of both `0' and both `1' to the initial gates of the circuit be desired, the circuit first has to be fully reversed, taking $t = 3$, and then executed in the forward direction with all inputs set to `0', taking nine units of time in total. If the system is to be returned to its original state, as is convenient in order to demonstrate thermodynamic reversiblility (see Section \ref{sec:irrevPipe}), a further three units of time will be required, for a total of twelve units in total.

\begin{figure}
	\includegraphics[width=\linewidth]{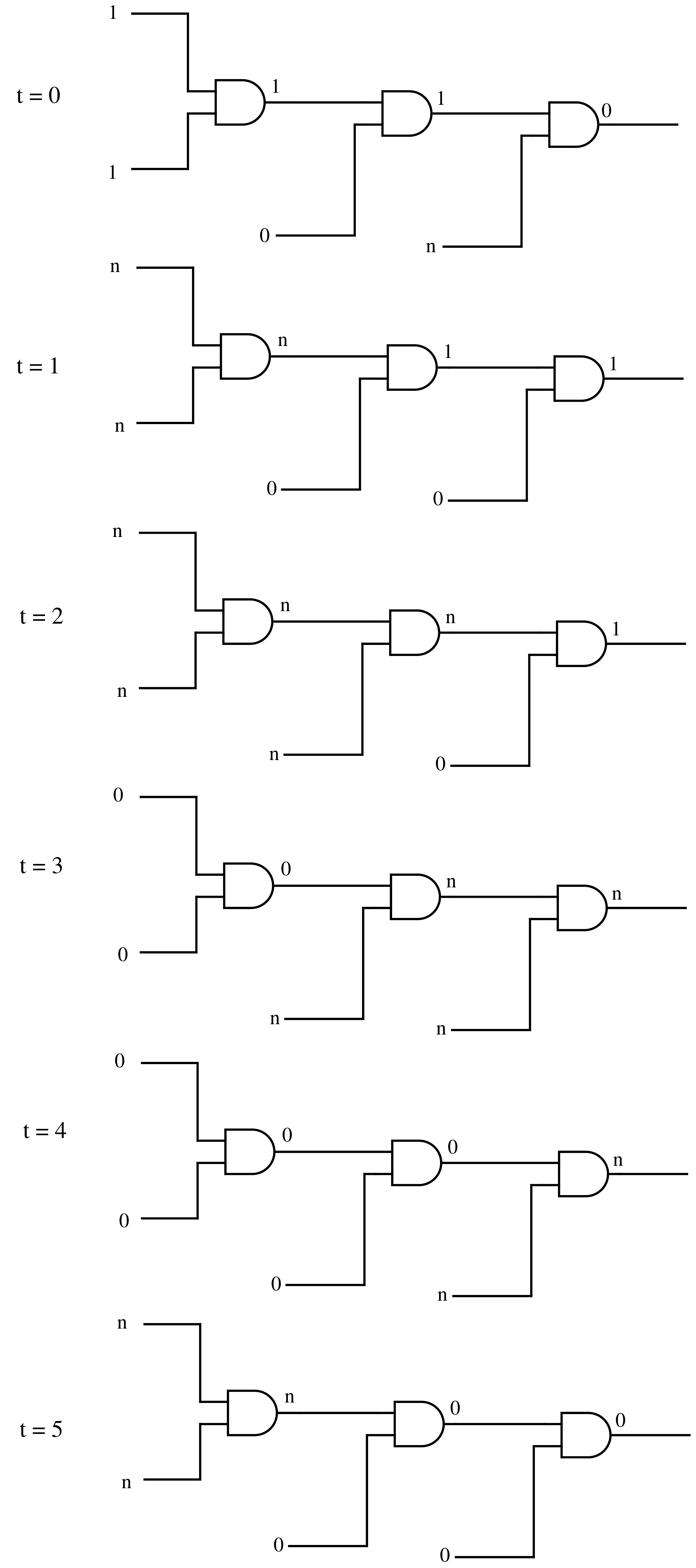}
	\caption{A simplified demonstration of pipelining in a circuit containing three chained AND gates. Units of time are arbitrary. States labelled `n' are neutral disengaged states set to neither `0' or `1'. The figure illustrates the five steps needed to compute two distinct outputs at the maximum possible throughput when the first input is delivered to the system at $t = 1$ and the second at $t = 4$. 
    The circuit becomes logically irreversible after $t =2$ due to the loss of information from the output of the final NAND gate.}
	\label{fig:pipelining}
\end{figure}

In contrast, for a pipelined system, information is passed through the circuit without waiting for all levels of logic gates to receive an input. In the case of the circuit depicted in Fig.\ \ref{fig:pipelining}, the `0' inputs are fed into the initial gate of the circuit before the result of the `1' inputs have been fully processed by the final gate. This results in a reduction of the time requirement to compute the results of both operations to only five units. Furthermore, the speed advantage of pipelining relative to phased chaining grows as the length of the circuit and the number of potential inputs increases. We also note that the pipelining approach does not require independently controlled external dipoles for each level of the circuit, unlike the phase-chained method. In this sense, therefore, the pipelining approach has a much simpler control mechanism.

The increase in throughput and speed does come at the cost of potentially rendering the circuit logically irreversible.
An operation on a system is logically reversible if the initial state of the information-bearing degrees of freedom can be unambiguously inferred from the final state.
In the literature on the thermodynamics of computation, it is typical to think about logical irreversibility in terms of whether the input of a computation, whether a single logic gate or a complex computing device, can be inferred from its output \cite{wol19}. In this context, a NAND gate appears to be logically irreversible, since it is impossible to tell from an output of 1 whether the input is 00, 01 or 10. 

When building a concrete architecture, however, it is often more natural that the outputs of a logical operation are not written over the inputs, which are preserved. This input-preserving property applies to our molecular mechanical circuits. Instead, the output from the first NAND gate in Fig.\ \ref{fig:phasedChaining} is written to a separate bit, which is then used as an input for the subsequent layer. Whether or not the circuit operates in a  logically reversible manner then depends not on the logic of the NAND gate, but rather on whether the initial state of the output bit is known unambiguously after the operation is complete.

In the specific case of the phase-chained system shown in Fig.\ \ref{fig:phasedChaining}, gate outputs are prepared in a neutral $n$ state prior to the computation ($t=0$). Each gate is updated only once, and therefore after each gate operation the initial state of the output bit is well known -- the output bit was initially neutral.

By contrast, during pipelining, each output bit is updated multiple times. During these processes, it cannot in general be guaranteed that the pre-update state can be unambiguously inferred from the post-update state. For example, in Fig.\ \ref{fig:pipelining}, the state of the system at $t=2$ cannot be unmabiguously inferred from the state at $t=3$; 
the value of ``1" in the final output bit at $t=2$ can only be inferred at $t=3$ if the state of the inputs to the first layer at $t=0$ are known, but this information is no longer present in the system at $t=2$, let alone $t=3$.

\section{\label{sec:irrevPipe}Logical and Thermodynamic Irreversibility in Pipelined Processes}


In order to properly determine the thermodynamic efficiency of a pipelined process, it is most practical to run the process in both forward and reverse directions such that the system returns to its original state. This is due to the fact that the output state of a computation may have a different free energy from the starting state, and may consist of an ensemble of states rather than a single state.

\begin{figure}
	\includegraphics[width=1.03\linewidth]{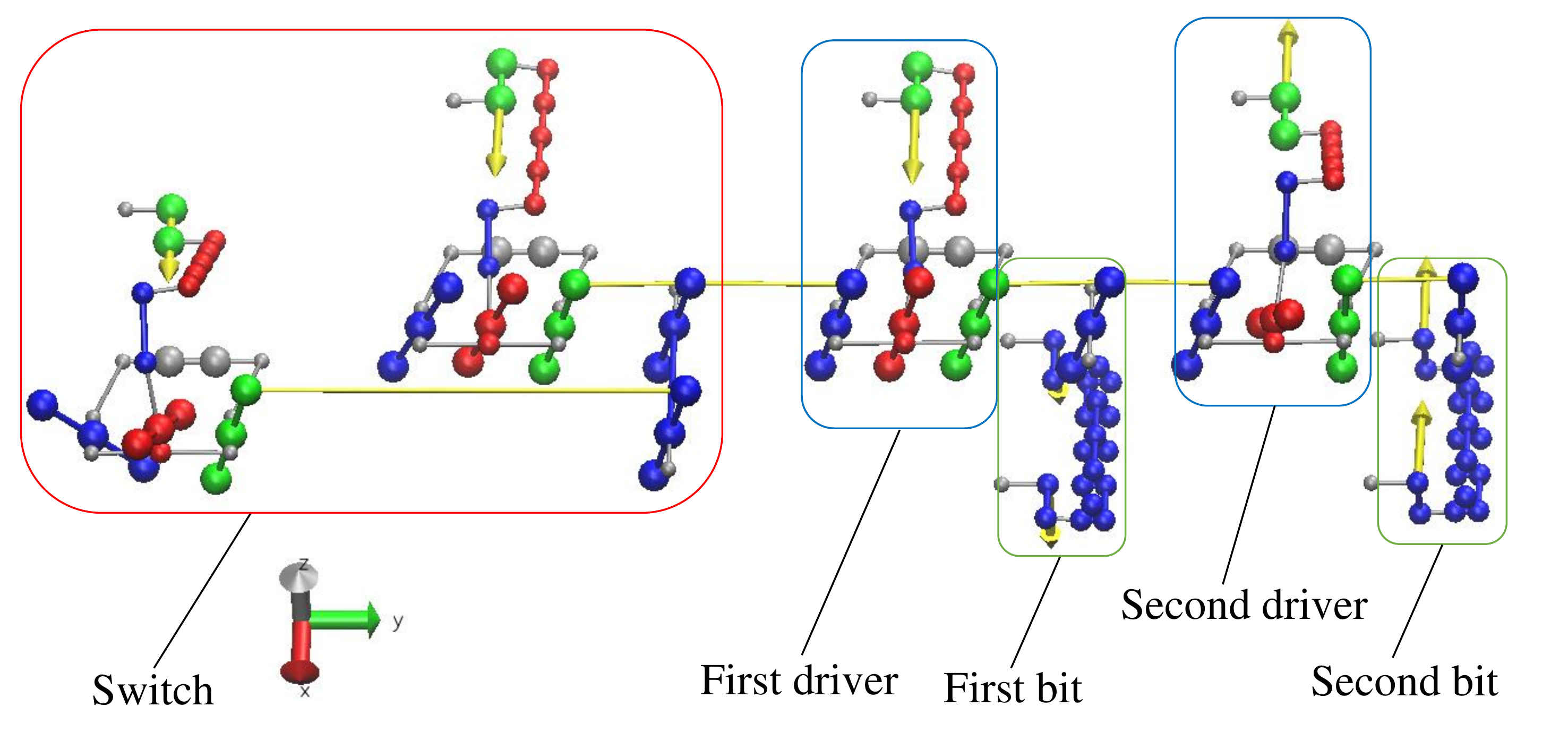}
	\caption{The two-bit pipelined system described in Section \ref{sec:irrevPipe} at $t^* = 0$.}
	\label{fig:pipeDemo}
\end{figure}
As previously mentioned, a pipelined circuit may become logically irreversible if its inputs are not stored beyond one clock cycle. This phenomenon can be demonstrated with a simple system, consisting of two bits chained to a switch, with each element being separated by a driver (Fig.\ \ref{fig:pipeDemo}). The two bits are equipped with signal boosters, which force the bit into either the `0' or `1' position when maximally engaged. The signal booster of the first bit is staggered 90\textdegree\ out-of-phase with the first driver while the signal booster of the second bit is in phase with the second driver and 90\textdegree\ out-of-phase with the signal booster of the first bit. The fact that the signal booster of the first bit is 90\textdegree\ out-of-phase with its driver allows the first bit to more effectively retain and transmit information to the driver of the second bit, since the information contained in the first bit is preserved for some time even after the driver of the first bit disengages. As mentioned in section \ref{ssec:switch}, the switch can be toggled between two different states.

In contrast to phased chaining, all internal dipoles of the system rotate at the same speed throughout the simulation regardless of initial starting position, with the exception of the dipoles of the switch which travel at only half the speed of the other dipoles due to the fact that the drivers of the switch only have to move through 180$\degree$ to engage with the other set of bits. In theory, both sets of dipoles can be driven by a single external dipole through the use of reduction gearing.

The two-bit system is designed to receive an input from the switch to the first bit and then transfer it to the second bit. The switch will be toggled between both possible inputs (`1' and `0'); this state will be transduced into the first bit, followed by the second bit after a short delay. As previously mentioned, these processes are driven by the rotation of an external dipole, which interacts with the internal dipoles in the system. After both inputs have reached the second bit, the dipoles driving the system will then be rotated in reverse for an identical duration. If the system were logically and thermodynamically reversible, every pair of points in the forward and reverse trajectories that are the same distance in time from the midpoint of the simulation should occupy the same ensemble of states. As we shall demonstrate, this is not the case for the pipelined two-bit system.

\begin{figure*}
	\includegraphics[width=1\linewidth]{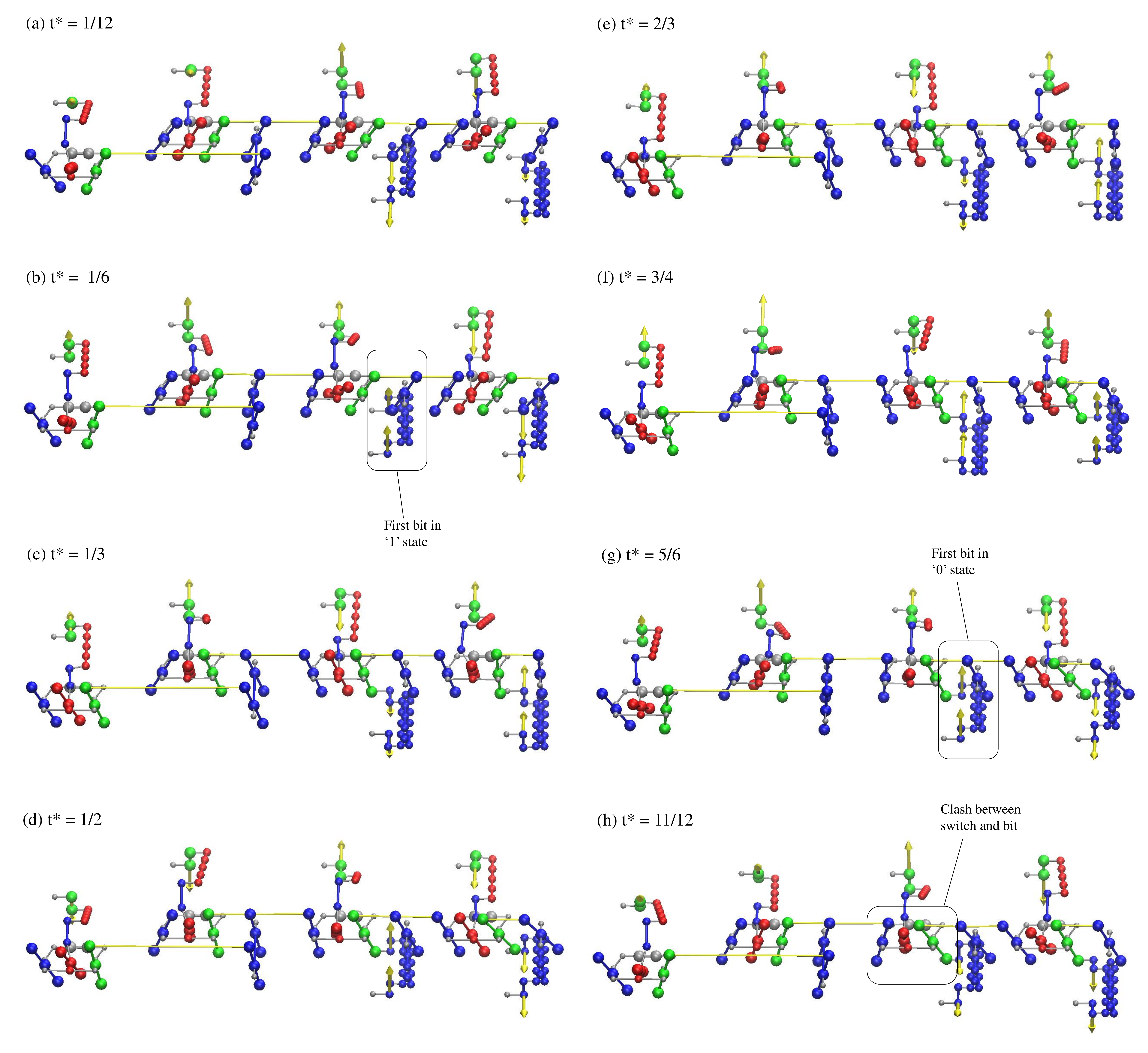}
	\caption{The two-bit pipelined system described in Section \ref{sec:irrevPipe} at various important values of $t^*$. Note the different states taken by the first and second bits in (a) and (h) as well as (b) and (g).}
	\label{fig:pipeDemoGallery}
\end{figure*}

At $t^* = 0$, the system is initialised with the signal boosters of both bits disengaged and the switch toggled to the `1' state. Here, $t^*$ is the ratio of the current time $t$ over the time required for the complete process that we are simulating.
From $t^* = 0$ to $t^* = 1/6$, the signal boosters of the first and second bits become engaged in turn, effectively copying the `1' state from the switch to the second bit. Between $t^* = 1/6$ and $t^* = 1/3$, the locks of the two bits disengage and the switch toggles between the `1' and `0' states. At $t^* = 1/3$, the switch is now toggled to the `0' state. From $t^* = 1/3$ to $t^* = 1/2$, the signal boosters of the first and second bits become engaged in turn, copying the `0' state from the switch to the second bit. The process is now reversed.

Between $t^* = 1/2$ and $t^* = 2/3$, information now flows from the second bit to the first bit. As the second bit was still locked into the `0' state at $t^* = 1/2$, the `0' state is correctly transduced to the first bit. However, a different situation develops at $t^* = 5/6$, when the signal booster of the second bit completes a 360\textdegree \ revolution. As the second bit was not constrained into any particular state before its signal booster is engaged, there is a chance it is forced into the `0' state instead of the `1' state that it was in at $t^* = 1/6$, a direct consequence of the logical irreversibility of the system. The chance of this occurring, henceforth referred to as the failure rate, is approximately 50\% at the highest clocks speeds and decreases with clock speed (Table \ref{tab:failureRate}), a decrease that is likely due to information leaking through the disengaged clutches to the bits due to imperfect clutching. This `0' state is transduced to the first bit at $t^* = 11/12$.

It is useful here to highlight the distinction between failure and error rate as defined in this paper. Error rate is defined as the ratio between the sum of simulations in which the logical output does not correspond to the expected result and the total number of simulations, and is thus applicable only to processes for which there is a clearly defined logical output (for instance, the states taken by the bits during the forward process). On the other hand, failure rate is defined as the ratio between the sum of simulations in which the state the system occupies at a given point in the reverse process is not identical to the state of the system at the corresponding point in the forward process, and the total number of simulations. 

\begin{figure}
	\includegraphics[width=\linewidth]{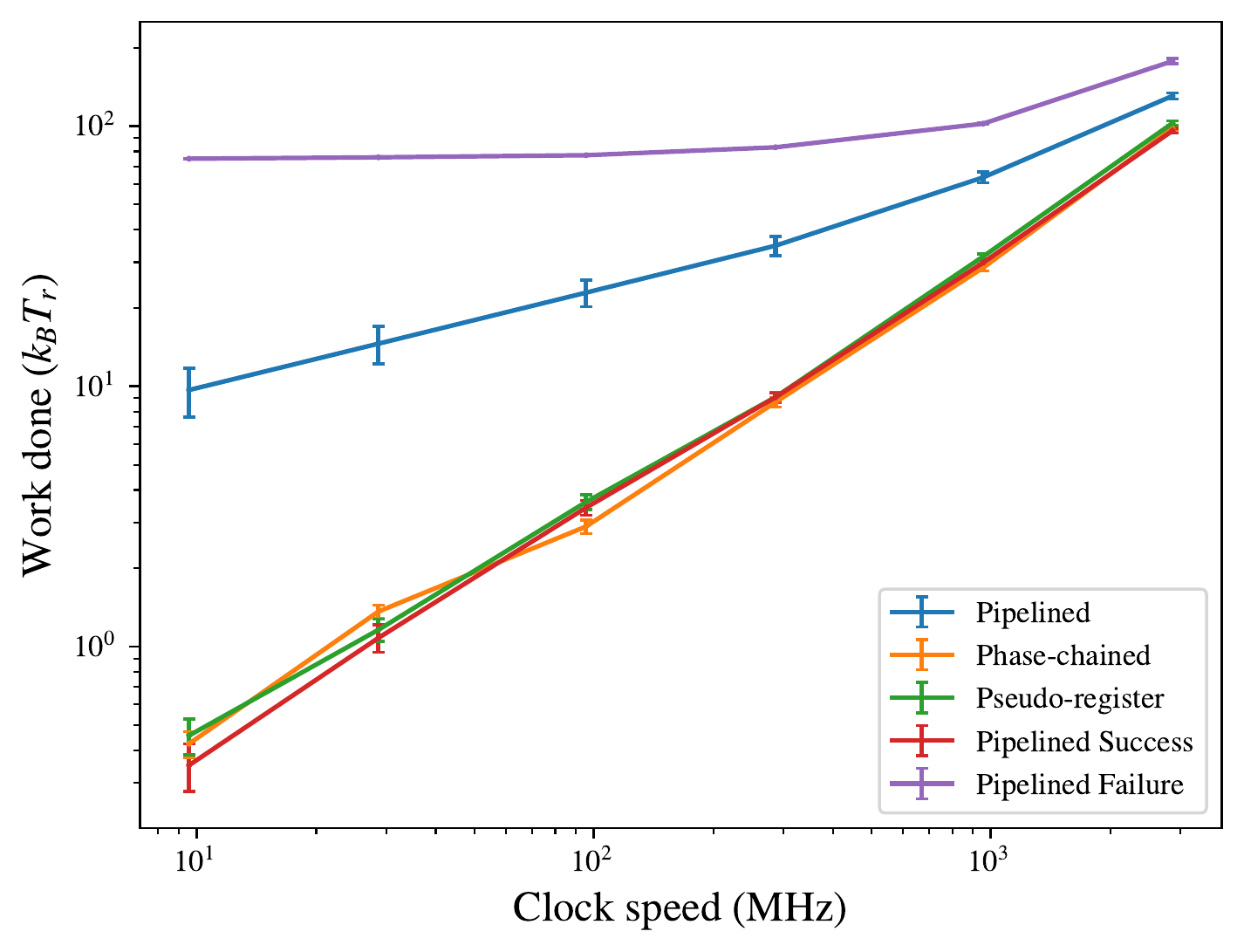}
	\caption{Work done against clock speed for the two-bit system 
 with the pipelining, phase-chaining and pseudo-register protocols, with separate plots for successful and failed reverse pipelining operations. 144 simulations were run for each data point. Before all protocols described in this paper were executed, the system involved was subject to energy minimisation followed by thermal equilibration. All simulations were run at 298\,K.}
	\label{fig:pipeDemoGraph}
\end{figure}

As the clutch of the driver located between the switch and the first bit engages and brings the two components into mechanical contact, it causes a clash between the first bit, forced into to the `0' state by its signal booster, and the switch, currently toggled to the `0' state. This clash is resolved by the first bit being forced into the `1' state despite the engagement of its signal booster, a process that is highly thermodynamically inefficient and results in the conversion of a large amount of potential energy into irrecoverable kinetic energy and subsequently heat, as can be observed from the results recorded in Fig.\ \ref{fig:pipeDemoGraph}. Regardless of clock speed, 
a failed reverse process requires approximately an additional 75\,$kT$ of work to complete compared to a successful reverse process. An animation of a failed reverse process can be viewed in Supplemental Video 1b, while an animation of a successful reverse process can be viewed in Supplemental Video 1a. The reduction in failure rate allows for a slight increase in thermodynamic efficiency as the clock speed is decreased, but the reduction in work done as a function of clock speed is still far less than the 1:1 ratio which would be expected if viscous drag were the primary cause of efficiency losses.

\begin{table}
\caption{Failure rate of the two-bit pipelined system relative to clock speed. A reverse pipelining process is defined as a failure if the second bit is in a `0' state rather than a `1' state at $t^* = 3/4$.}
\begin{tabular}{c@{\hspace{1cm}}c}
\midrule
\midrule
Clock speed (MHz) & Failure rate \\ 
\midrule
\centering
2870 & 0.493 \\ 
957 & 0.465 \\ 
287 & 0.347 \\ 
95.7 & 0.264 \\ 
28.7 & 0.181 \\ 
9.57 & 0.125 \\ 
\midrule
\midrule
\end{tabular}
\label{tab:failureRate}
\end{table}


Notably, this clash does not occur when the system is driven via phased-chaining, which is both logically and thermodynamically reversible as the signal booster of the second bit is never engaged when the signal booster of the first bit is disengaged. 

As can be observed from Fig. \ref{fig:pipeDemoGraph}, phased chaining is comparable in thermodynamic efficiency to a successful pipelining operation, and is thermodynamically reversible. An animation of the two-bit system driven via phased chaining is available in Supplemental Video 1c.

\subsection{The pseudo-register}

\begin{figure*}
	\includegraphics[width=0.75\linewidth]{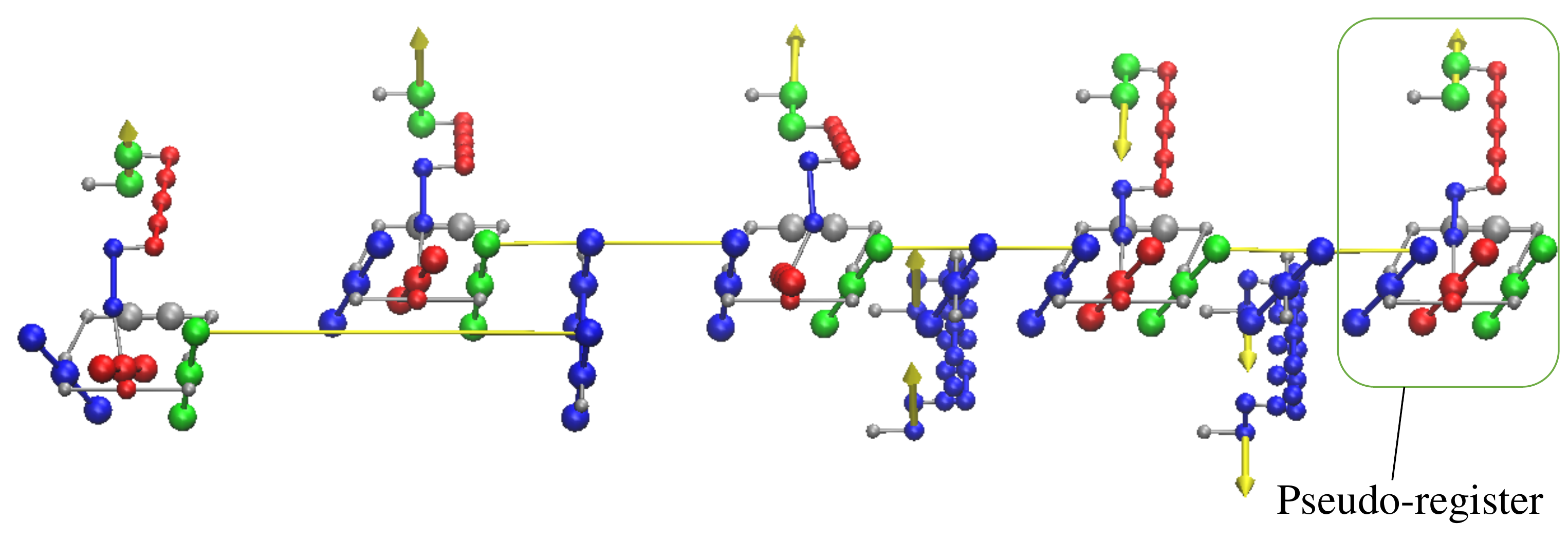}
	\caption{The two-bit pipelined system chained to a pseudo-register at $t^* = 5/6$.}
	\label{fig:pseudo}
\end{figure*}

Mitigating the loss of efficiency associated with logical irreversibility is possible if the circuit is rendered logically reversible by preserving the information contained in the second bit at $t^* = 3/4$. In the absence of sequential logic, this feat can be artificially accomplished by the chaining of an additional driver set to `1' to the second bit (Fig.\ \ref{fig:pseudo}). The dipole of this driver was synchronised to the signal booster of the second bit such that the driver returns `1' at the same time that the signal booster of the second bit reaches its maximal engagement. Thus, this driver will invariably force the second bit into the state `1'. As this driver simulates the information from the output bit being stored in a register even after the switches are no longer engaged before being subsequently re-written into the output bit during the reverse process, it shall henceforth be referred to as a pseudo-register. An animation of the two-bit pipelined system chained to a pseudo-register can be viewed in Supplemental Video 2.

The results of the simulations with the switch-linked two-bit pipelined system chained to the pseudo-register are summarised in Fig.\ \ref{fig:pipeDemoGraph}. From these results, it can be observed that preserving the input and thus rendering the circuit logically reversible removes the thermodynamic costs associated with pipelining in this case.
\newline

\section{\label{sec:revPipe}Pipelining in larger circuits}
\begin{figure}
	\includegraphics[width=\linewidth]{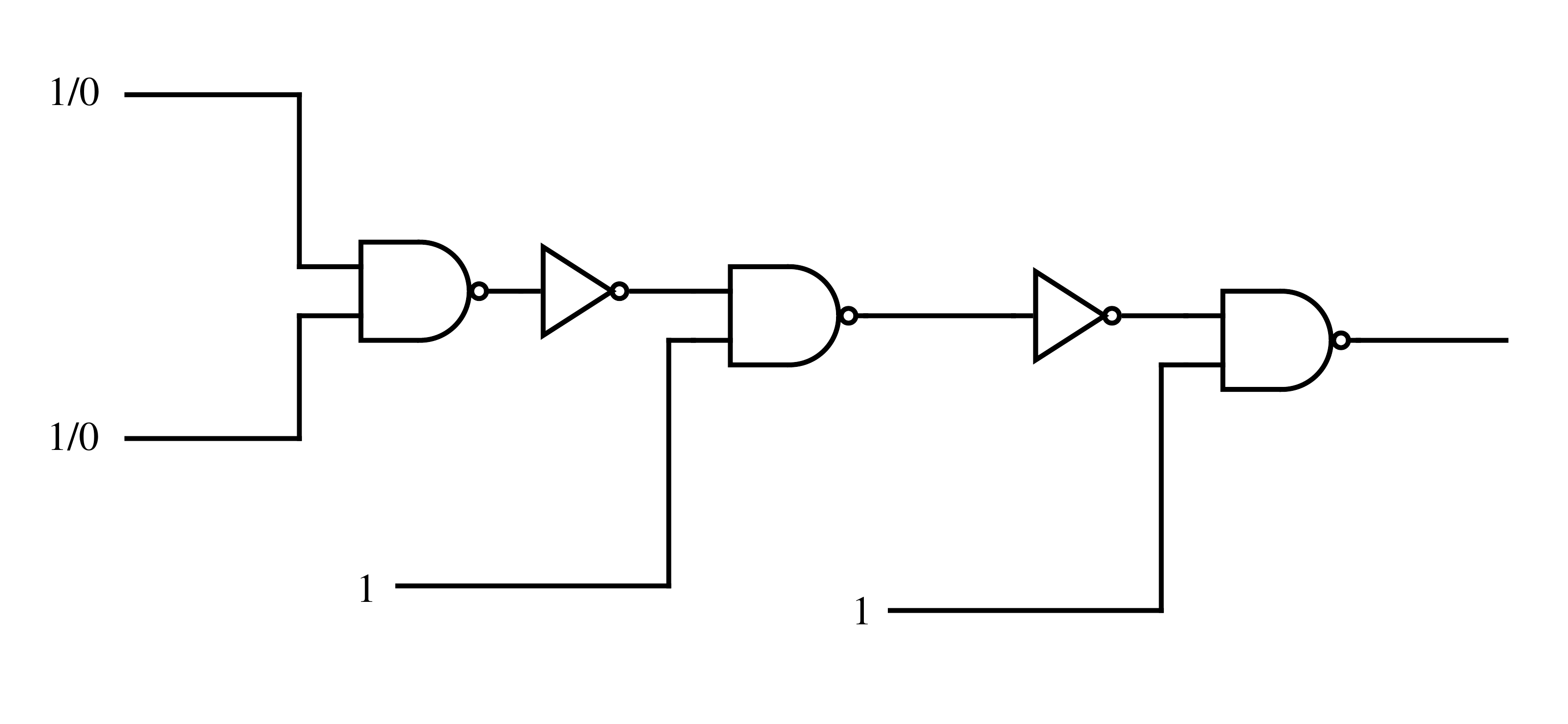}
	\caption{One of three circuits used to demonstrate reversible pipelining. This particular circuit has a chain length of 3 and can be extended or truncated by the addition or removal of inverter-NAND units. The other two circuits have lengths of 2 and 4 NAND gates. This particular circuit is equivalent to the circuit outlined in Fig.\ \ref{fig:pipelining} since the NAND-inverter units are equivalent to an AND gate.}
	\label{fig:pipeCircuit}
\end{figure}

A series of circuits equivalent to the set used to demonstrate phased chaining in our previous work \cite{seet22} was used to implement a larger pipelined circuit (Fig.\ \ref{fig:pipeCircuit}). From the standpoint of ensuring logical reversibility, this circuit has the useful feature of being conditionally invertible; that is, it is possible to fully reconstruct the input of a given NAND gate from its output as long as the other input bit (in this case, `1') is preserved. However, as mentinoed in Section \ref{ssec:phasedChain}, the circuit in Fig.\ \ref{fig:pipeCircuit} does not have a pseudo-register attached to preserve the final input bit and is therefore logically irreversible in its current form.

\begin{figure}
	\includegraphics[width=\linewidth]{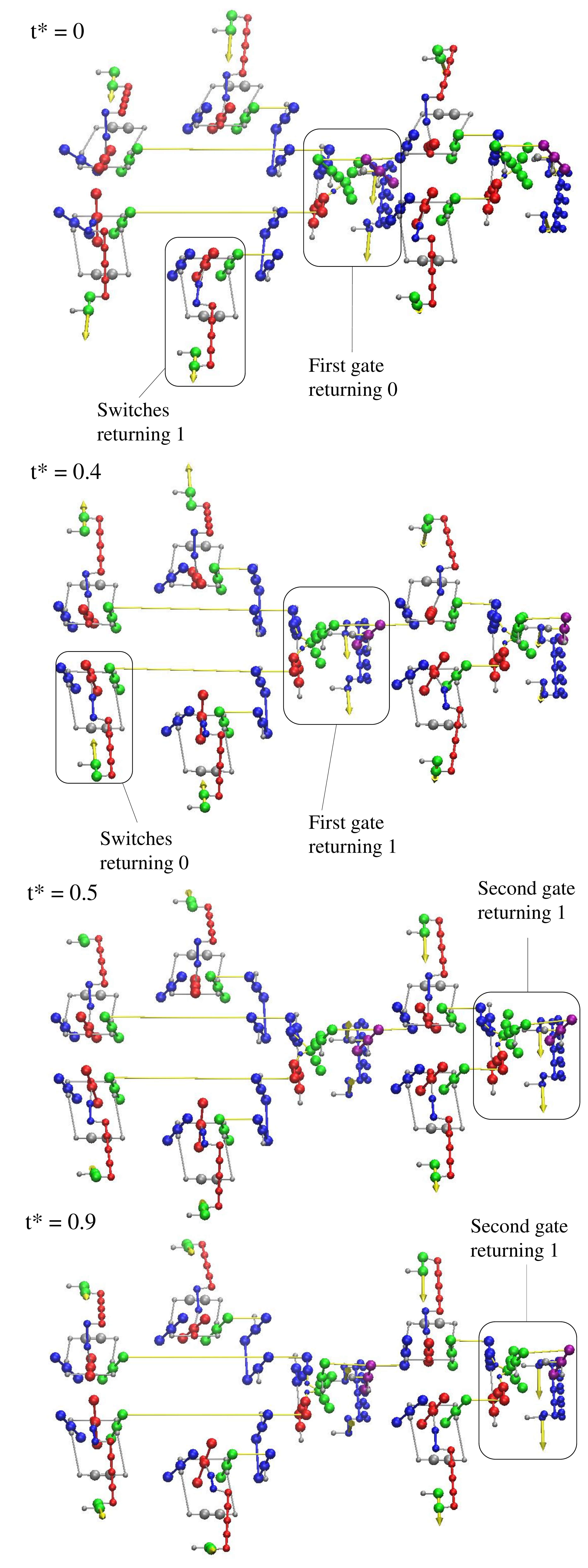}
	\caption{The two-gate pipelined system at four critical values of $t^*$.}
	\label{fig:pipeGallery}
\end{figure}
In a manner analogous to the two-bit pipelined system, the phases of each gate are offset by 90$\degree$ such that the signal booster of any given gate begins to engage when the signal booster of the previous gate becomes maximally engaged. The protocol used is equivalent to that described in Fig.\ \ref{fig:pipelining} since each NAND-inverter unit is equivalent to an AND gate.

To better illustrate the details of the implementation of this specific pipelining protocol, we will use the example of two chained NAND gates (Fig.\ \ref{fig:pipeGallery}). At $t^* = 0$, the dipoles of the switches are initialised such that the 
switches return `1' and the signal booster of the first gate is initialised at the fully engaged position. The drivers and signal boosters of the second gate are initialised 90$\degree$ behind the signal booster of the first gate. It is at this point that the first-phase error rate of the first gate is measured. At $t^* = 0.1$, the drivers and signal booster of the second gate become fully engaged and the first-phase error rate of the second gate is measured. At $t^* = 0.4$, the switches have fully transitioned to the `0' state and the signal booster of the first gate is fully engaged, returning `1'.  It is at this point that the second-phase error rate of the first gate is measured. At $t^* = 0.5$, the drivers and signal booster of the second gate become fully engaged and the second-phase error rate of the second gate is measured.

At $t^* = 0.5$, the dipoles driving the circuit are reversed. During this reverse process, the drivers and signal boosters of the second gate are maximally engaged at $t^* = 0.5$ and $t^* = 0.9$, while the signal booster of the first gate is maximally engaged at $t^* = 0.6$ and $t^* = 1$. However, at $t^* = 0.9$, the first gate has already been disengaged for three quarters of a clock cycle and no longer carries any useful information. Therefore, as the signal booster of the second gate engages, it falls on a randomised rather than a well-defined state. This loss of saved information results in the circuit becoming logically irreversible, as was the case with the two-bit pipelined system in the absence of a pseudo-register. Animations of for the two, three and four NAND gate pipelined circuits can be viewed in Supplemental Videos 3-1, 3-2 and 3-3, respectively.

\begin{figure}
	\includegraphics[width=1\linewidth]{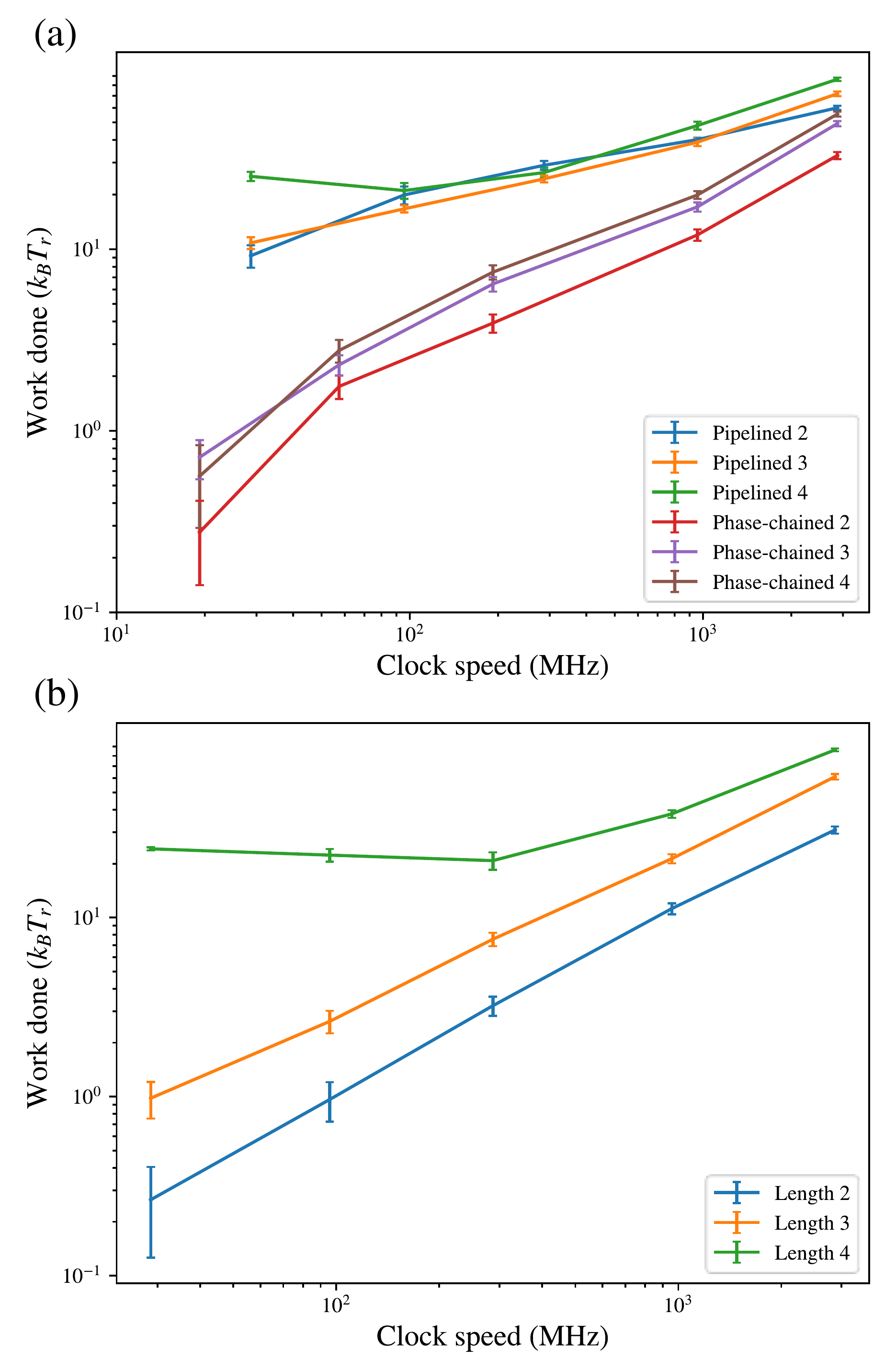}
	\caption{(a) Work done against clock speed for different chain lengths of the NAND gate circuit driven via the pipelining and phased chaining protocols. (b) Work done against clock speed for different chain lengths of the NAND gate circuit driven via the pipelining protocol where the final NAND gate is chained to a pseudo-register. Each data point corresponds to an average over 144 simulations.}
	\label{fig:mpd}
\end{figure}

Simulations to determine the thermodynamic efficiency for the pipelined NAND circuit were executed according to the protocol described in the previous paragraphs.
The results are summarised in Fig.\ \ref{fig:mpd}(a). The work done in the pipelining processes is an order of magnitude higher than that of the analogous system driven via phased chaining described in Ref.\ \cite{seet22}. Failure rates for the reverse pipelining processes for the pipelined NAND circuit and the work done for successful and failed reverse pipelining processes are recorded in the Supplementary Material. Unlike the two-bit pipelined system, the differences in thermodynamic efficiency between successful and failed reverse pipelining processes, while still present, are less clear, and the successful pipelining processes are still significantly less efficient than their phase-chained counterparts. The lack of a clear trend is possibly due to the greater complexity of the system, with factors such as the error rate of the pipelining process affecting the failure rate.

\begin{figure*}
	\includegraphics[width=1\linewidth]{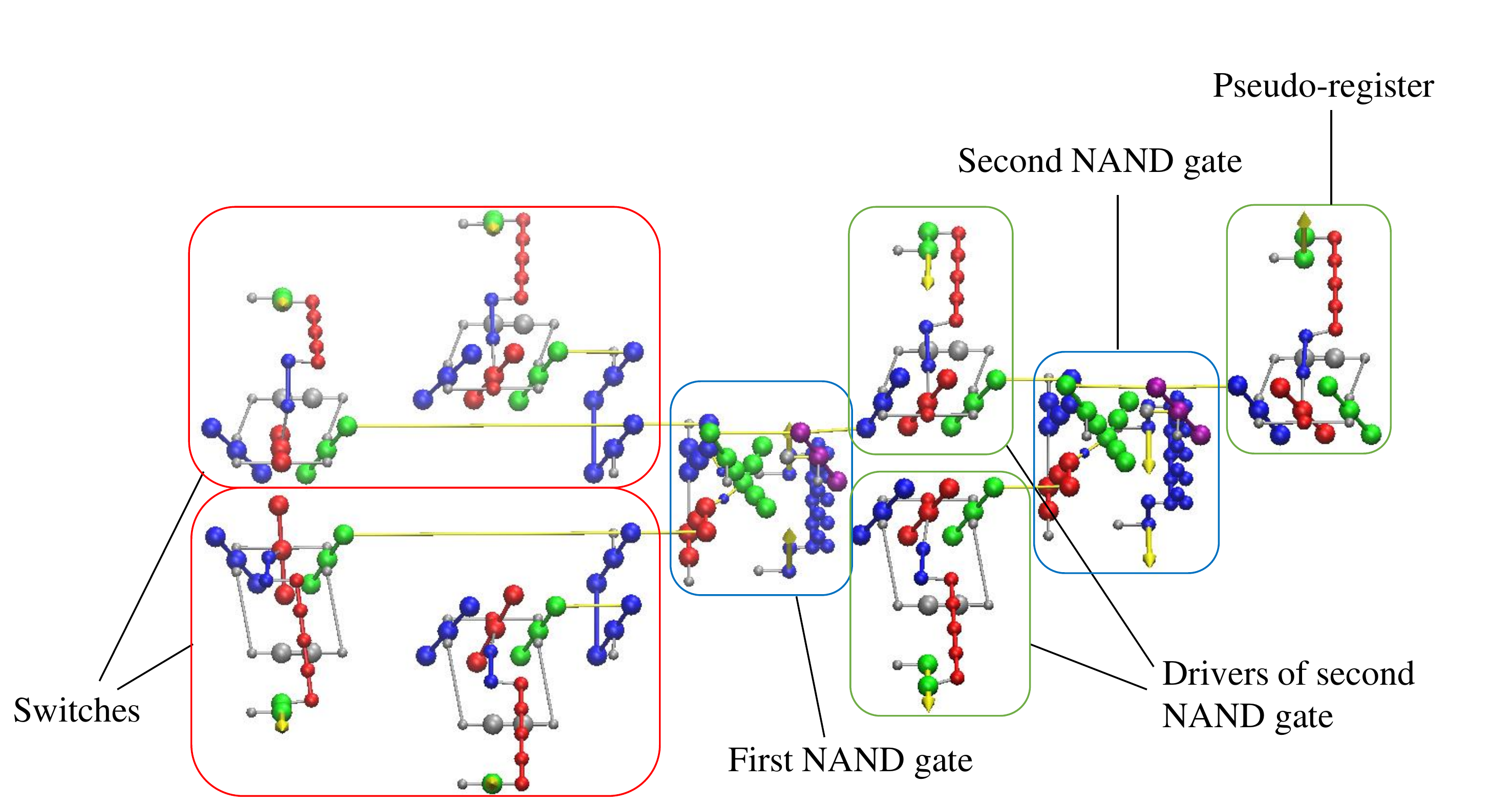}
	\caption{Illustration of the system used to test the pipelining protocol with two gates, modified such that the final gate is chained to a pseudo-register.}
	\label{fig:pipeDriverLabel}
\end{figure*}

In an attempt to render the circuit logically reversible, a pseudo-register set to `0' was chained to the end of the final NAND gate, as illustrated in Fig.\ \ref{fig:pipeDriverLabel}. Such a system should, in theory, be logically reversible. Animations of for the two-, three- and four-NAND gate pipelined circuits chained to a pseudo-register can be viewed in Supplemental Videos 4-1, 4-2 and 4-3a respectively. The results of the reversibility simulations were run for this modified system are summarised in Fig.\ \ref{fig:mpd}(b). While this modification does restore approximate thermodynamic reversibility in the two- and three-gate systems, results for the four-gate system do not differ significantly from the unmodified example.

The failure rate of reverse pipelining in pseudo-register equipped circuits was not detectable over 144 simulations regardless of length or clock speed; thus, failed reverse pipelining cannot be the reason the four-gate pipelined system is much less efficient than its counterparts. Instead, the most likely reason for the thermodynamic irreversibility of the four-gate system is its high error rate, as can be observed from the results of error-rate simulations performed on the system with an additional pseudo-register 
in Table \ref{tab:pipeErr} in the forward pipelining process. Relative to phased-chained circuits of the same type, pipelined circuits generally have more simultaneous moving parts, increasing the difficulty of optimisation to reduce error rate. In this particular case, the high error rate is likely due to the inability of the clutch to fully prevent the upstream gates from influencing the downstream gates even when fully disengaged; an animation of this erroneous state can be viewed in Supplemental Video 4-3b. The thermodynamic irreversibility induced by the high error rate highlights the importance of optimising the clutching mechanism of the driver if effective pipelining is desired.

\begin{table}
\caption{Results obtained from error-rate simulations at $T_r$ (298\,K) of the pseudo-register equipped two, three and four chained NAND gates driven via the pipelining protocol at 2.87\,GHz. 6000 simulations were run for each data point. Similar results are obtained in the absence of a pseudo-register.}
\begin{ruledtabular}
\begin{tabular}{cccc}
Chain & Gate & 1\textsuperscript{st} phase & 2\textsuperscript{nd} phase \\
Length & number & error & error \\
\midrule
\centering
2 & 1 & 0 & 0 \\
2 & 2 & 0 & 0 \\
3 & 1 & 0 & 0 \\
3 & 2 & 0 & 0 \\
3 & 3 & 0 & $5.11 \times 10^{-3}$ \\
4 & 1 & 0 & 0 \\
4 & 2 & 0 & 0 \\
4 & 3 & 0 & $4.00 \times 10^{-4}$\\
4 & 4 & 0 & 0.0339
\end{tabular}
\end{ruledtabular}
\label{tab:pipeErr}
\end{table}

\subsection{The modified driver}
\begin{figure}
	\includegraphics[width=1\linewidth]{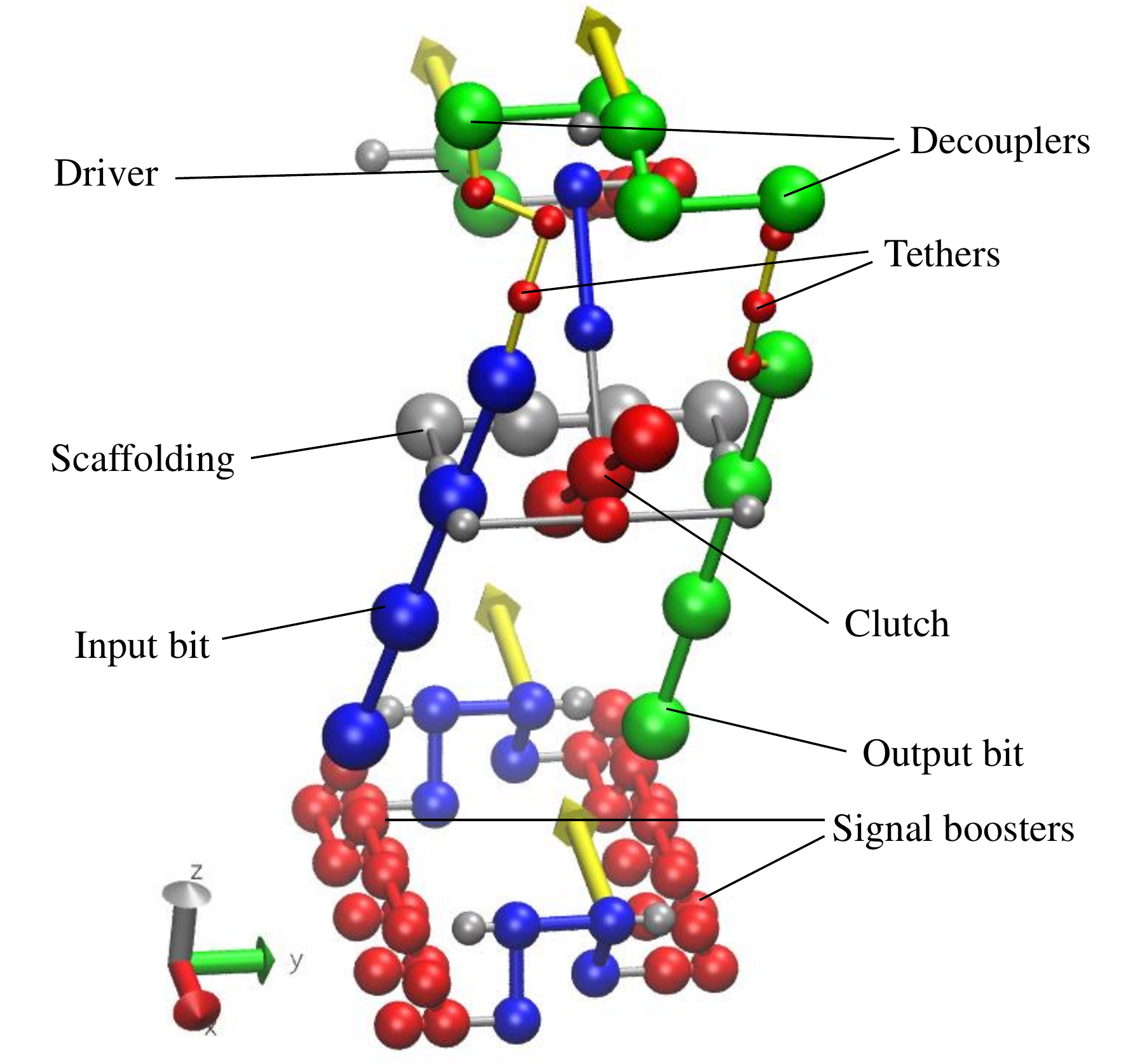}
	\caption{The modified driver used to lower the error rate in order to demonstrate reversible pipelining.}
	\label{fig:dnbLabel}
\end{figure}

In order to reduce the error rate of the pipelined circuit, a new driver was designed with a different clutch mechanism (Fig.\ \ref{fig:dnbLabel}). The new driver design incorporates two major modifications: a pair of decouplers which are connected via tethers to the input and output bits, and a pair of signal boosters similar to those used in the NAND gates. The purpose of the decouplers is to prevent information from the upstream bits from leaking through to the downstream bits by applying a pulling force through the tethers (coloured red in Fig.\ \ref{fig:dnbLabel}), forcing the bit to which it is connected into a neutral position. Decouplers and signal boosters are required for both the input and output bits due to the fact that the circuit must be operated in both forward and reverse in order to demonstrate reversibility. The decouplers for the input and output bits are offset by 120\textdegree \ and the signal boosters by 60\textdegree.

\begin{figure*}
	\includegraphics[width=1\linewidth]{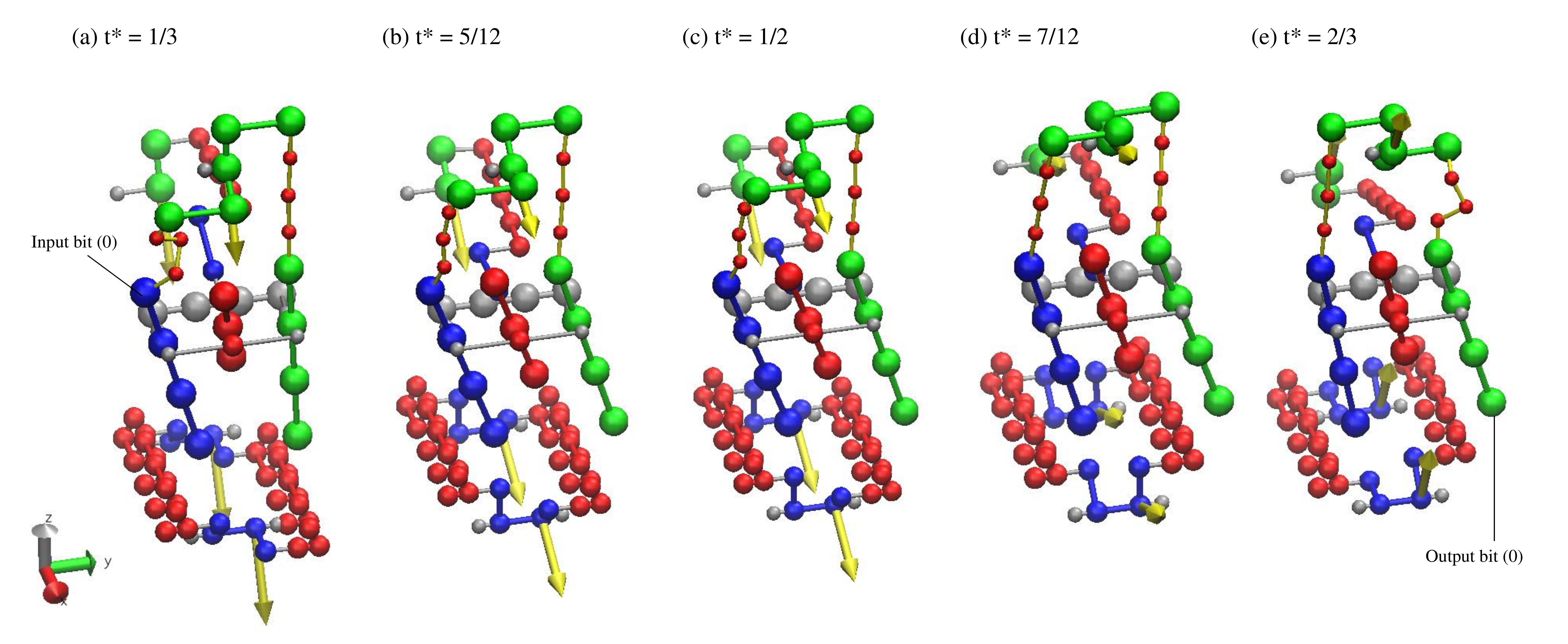}
	\caption{The modified driver at five critical values of $t^*$ during the process of transducing the state `0' from the input bit to the output bit. 
    }
	\label{fig:dnbGallery}
\end{figure*}

Fig.\ \ref{fig:dnbGallery} illustrates the operation of the modified driver. In this example, the values of $t^*$ correspond to fractions of exactly one clock cycle. Intuitively, in order to maximise the probability of error-free information transduction, the point at which the driver reaches maximum engagement ($t^*$=0.5) should correspond to the midpoint of the phases of the gates chained to its input and output bits. Therefore, given the 90\textdegree \ phase difference between an upstream and downstream gate, the peak engagement of the upstream gate occurs 45\textdegree \ before the peak engagement of the driver at $t^* = 3/8$; correspondingly, the downstream gate reaches peak engagement 45\textdegree \ after the driver does so at $t^* = 5/8$.

Before the upstream gate reaches peak engagement, the decoupler for the output bit engages at $t^* = 1/3$. This forces the output bit into an approximately neutral state and thus improves the ability of the clutch to prevent information from flowing from the downstream gate to the upstream gate or vice versa when not desired. As previously stated, the upstream gate reaches peak engagement at $t^* = 3/8$, followed by the signal booster of the input bit at $t^* = 5/12$, the driver itself at $t^* = 1/2$. By this time the decoupler for the output bit has become sufficiently disengaged that it no longer exerts a significant force on the bit, allowing the signal booster of the output bit to engage fully at $t^* = 7/12$, finally transducing information to the downstream gate which reaches peak engagement at $t^* = 5/8$. This process allows information to flow smoothly from the upstream gate to the downstream gate. At $t^* = 2/3$, the decoupler for the input bit becomes fully engaged. This process, which grants the driver symmetry regardless of the direction the external dipole rotates, is unnecessary for the transduction of information in the forward direction; however, it is a critical component of the reversible pipelining process.

\begin{figure}
	\includegraphics[width=1.05\linewidth]{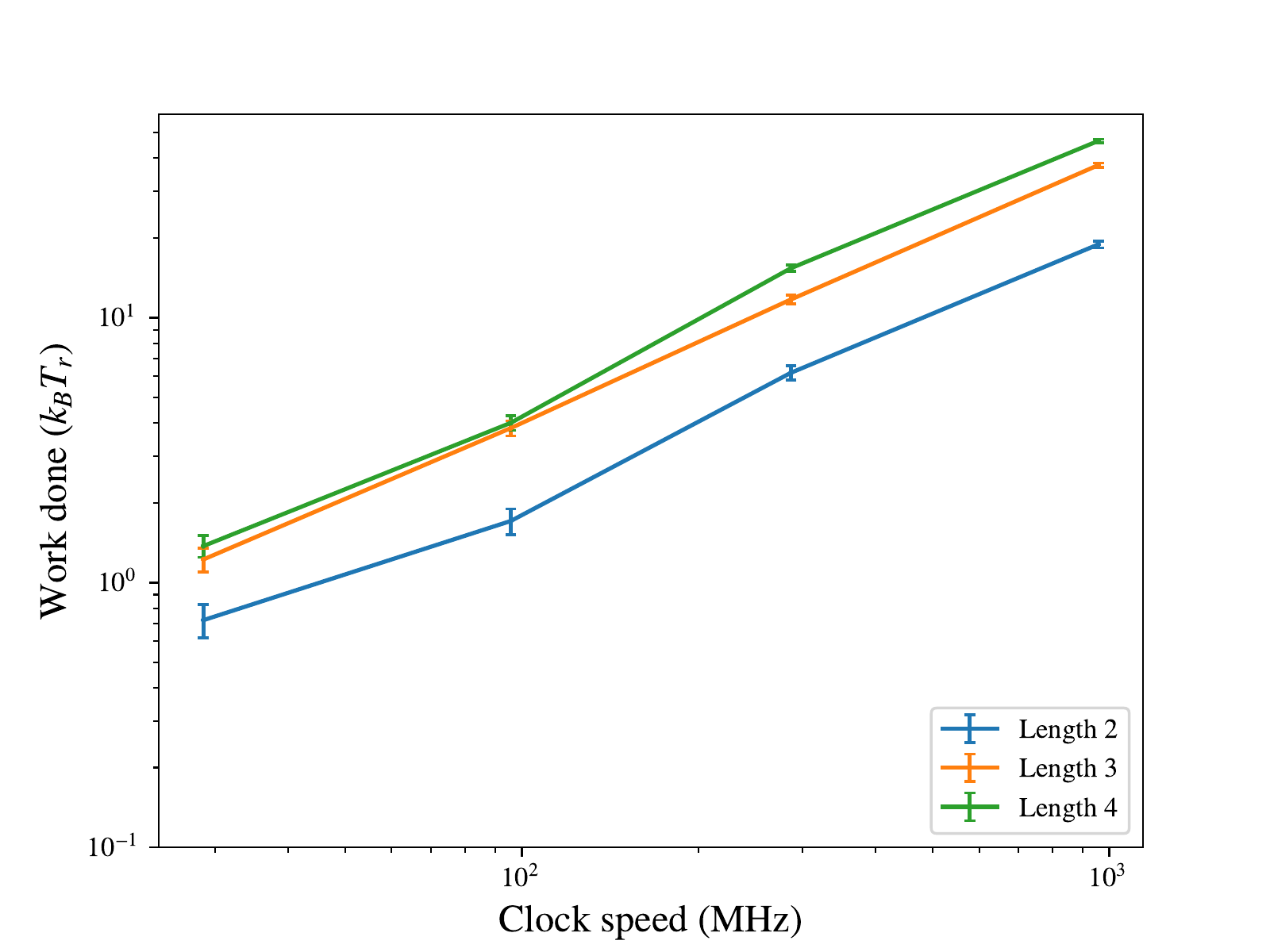}
	\caption{Plots of work done against clock speed for the pipelined system with two, three and four chained NAND gates with the modified driver. Each data point corresponds to 96 simulations.}
	\label{fig:dipe3Graph}
\end{figure}

A figure illustrating the four-gate pipelined system with the modified drivers replacing all drivers that connect two NAND gates is available in the supplemental materials. Drivers that connect fixed input bits to gates and the pseudo-register have been left unchanged. An animation of the four-NAND gate pipelined system with modified drivers can be viewed in Supplemental Video 5. The results of error rate and reversibility simulations run on the system are recorded in Table\ \ref{tab:dipe3Err} and Fig.\ \ref{fig:dipe3Graph}. Based on the high temperature needed to induce errors, it can be concluded that the error rate of the system with modified drivers is far lower than that of the unmodified system at room temperature. In addition, as the work done by the external dipoles of the modified system is approximately proportional to clock speed for all chain lengths, it can be concluded that this system converges toward thermodynamic reversibility in the limit of clock speeds tested.        

\begin{table}
\caption{\label{tab:dipe3Err}Results obtained from error-rate simulations at $6T_{r}$ (1788\,K) of four chained NAND gates connected by modified drivers at 2.87\,GHz. 1200 simulations were run for each data point. No errors were detected at simulations run at $T \leq 3T_{r}$}
\begin{ruledtabular}
\begin{tabular}{ccc}
Gate number & 1\textsuperscript{st} phase error & 2\textsuperscript{nd} phase error \\
\midrule
\centering
0& $7.50 \times 10^{-3}$& 0 \\
1& $2.50 \times 10^{-3}$& $1.67 \times 10^{-3}$ \\
2& $3.33 \times 10^{-3}$& $7.50 \times 10^{-3}$ \\
3& $5.83 \times 10^{-3}$& $1.00 \times 10^{-2}$ \\
\end{tabular}
\end{ruledtabular}
\end{table}

\section{Conclusion}
Pipelining provides a method of greatly increasing the throughput and reducing the number of external controls of a combinatorial circuit; however, these advantages are offset by a number of limitations. We have demonstrated the thermodynamic inefficiency in pipelined circuits by comparing the thermodynamic efficiency of the same circuit driven via pipelining and phased chaining. Moreover, pipelined circuits are more vulnerable to erroneous computational outputs, which themselves contribute to the thermodynamic inefficiency. 

It is interesting to note that low accuracy is associated with additional costs in this realisation of a logical circuit. In the field of the thermodynamics of computation, it is common to think of reduced accuracy as a way to mitigate computational cost \cite{ould17, rie20}.
This intuition fails here because the reduction in accuracy comes not from pushing the system less hard towards the desired end state with the external control, as is typically the case, but because the pushes coincide with configurations that are poorly aligned with the device's intended operation, causing clashes that are highly unfavourable. 

The inaccuracy of the computation in pipelined systems is, itself, a source of logical irreversibility,
and the clashes that occur are similar to the clashes that occur due to the logical irreversibility that is inherent to a pipelined set-up.
Rather than try to differentiate between logical irreversibility and inaccuracy as a source of thermodynamic costs, we instead point to the fact that the overall costs are quite large relative to what might be expected from a mismatch cost for the logical operation, based on a direct application of the abstract theory in Ref.~\cite{wol19,wol20}.

Recall that for any protocol that generates logically irreversible behaviour of a system, there is an optimal initial distribution $q(x)$ over the possible states $x$ of the system that minimizes entropy production for that protocol \cite{wol19, wol20}. For an arbitrary input distribution $p(x)$, the entropy production is split into two parts: a ``mismatch cost" that is upper-bounded by $k_{\rm B}T D(p(x)||q(x))$, and a ``residual entropy production''. $D(p(x)||q(x))$  is the K\"{u}llback-Leibler divergence between $p(x)$ and $q(x)$; the mismatch cost is therefore related to how far the actual input distribution $p(x)$ differs from the optimal distribution for the protocol, $q(x)$. The residual entropy production is related to the physical details of how the computation is performed, rather than just the computational logic of the input-to-output map and in principle can typically be taken to zero if the operations are performed quasistatically.

Although we have not attempted to evaluate $q(x)$ in our setting, it is unlikely to be particularly strongly biased towards any given logical input. For an unbiased $q(x)$, $D(p(x)||q(x))$ is upper bounded by $\ln 2$ per bit in a truly discrete logical system, whereas we observe larger costs than this. We now consider how these larger observed costs can, in principle, be reconciled with this abstract theory.

One possible explanation for the difference between $k_{\rm B}T D(p(x)||q(x))$ and the observed entropy production is that there is a substantial residual entropy production. However, although this claim cannot be confirmed without knowledge of $q(x)$, we suspect that the residual entropy production for our system is low in the limit of slow clock cycles, since the same gates converge on thermodynamic reversibility when operated in a phase-chained manner at slow operation speeds.

An alternative possibility is that the extra cost can be justified in terms of the breakdown of the digital logic approximation; for instance, when the signal boosters are in the process of disengaging and there is a surmountable energy barrier between the `0' and `1' states. The molecular mechanical system does not have truly discrete states, but rather wells within a continuous and high-dimensional energy landscape. 
The distributions $p(x)$ and $q(x)$ should, in principle, be defined over this continuous state space, rather than the discrete approximation. In this setting, the upper bound of  $k_{\rm B}T \ln 2$ per bit on the K\"{u}llback-Leibler divergence does not strictly apply. Although it is not immediately clear how this effect of continuous degrees of freedom feeds through into the entropy production, it is interesting to note that the most costly trajectories are associated with clashes in which the digital originlogic has not functioned as planned.

Regardless of the 
actual origin of the observed thermodynamic costs in this particular system, our simulations have highlighted a concept that applies more generally to systems in which discrete logical states are approximated by continuous degrees of freedom. Specifically, information theoretic bounds applied with an assumption of discrete states may be misleading, particularly when the assumed digital logic does not function as intended. Similarly, it is reasonable to assume that the reduced accuracy of pipelining-based methods probably applies beyond the specific setting considered here, although simulating the gates with an explicit model was helpful in demonstrating this phenomenon.

Significant latitude remains for the further exploration of the thermodynamics of pipelined molecular mechanical circuits and the improvement of existing designs. Perhaps the most natural question to ask is: how much of the entropy production for round-trip operations is incurred on the forwards part of the trajectory, in which the computation is actually performed, and how much during the reversed part of the protocol? Based on the nature of the clashes observed, we speculate that much of the entropy production may occur during the reverse part of the protocol. However, testing this hypothesis would require a substantially more sophisticated analysis of the system, such as calculating the change in free energy alongside the work done during the computation, since the work done over and above the free energy change determines the entropy production \cite{esp11}. Alternatively, persistent information storage could potentially be used to render non-invertible circuits into fully logically reversible circuits at the cost of space. Further improvements to the clutching mechanism of the driver may be possible via the application of optimisation meta-heuristics such as a genetic algorithm.

\section{Additional Materials}
VMD-viewable trajectory files of the pipelined circuit of length up to four NAND gates and the modified driver are available in the Oxford Research Archive at \url{https://ora.ox.ac.uk/objects/uuid:76d6d630-ad63-4013-b06a-3e1567960afc}. Videos of the various circuits described in this paper are available at \url{https://www.youtube.com/playlist?list=PLnguwFTyIfT4ZG4Z9u024ywSxFubXukyI}.

\begin{acknowledgments}
The authors are grateful to the UK Materials and Molecular Modelling Hub and to the University of Oxford Advanced Research Computing (ARC) facility for computational resources. TEO is funded by a Royal Society University Research Fellowship.
\end{acknowledgments}

\bibliography{apssamp}

\include{sm.tex}
\end{document}

%% file: sm.tex
\section{Supplemental Material}
\setcounter{figure}{0} \renewcommand{\thefigure}{S\arabic{figure}} 
\setcounter{table}{0} \renewcommand{\thetable}{S\arabic{table}}



\subsection{Failure rate data for the pipelined NAND circuit}
\begin{table}[h]
\caption{Failure rate of the pipelined NAND circuits of length $n=$2, 3 and 4 relative to clock speed. A reverse pipelining process is defined as a failure if the last NAND gate in the chain is forced into `1' state by its signal booster during the second phase of the reverse pipelining process. This is at $t^*$=9/10, 10/12 and 11/14 for $n$=2, 3 and 4, respectively.}
\begin{tabular}{c@{\hspace{1cm}}ccc}
\midrule
\midrule
 Clock speed & & Failure rate & \\ 
\cline{2-4}
(MHz) & $n=2$ & $n=3$ & $n=4$ \\ 
\midrule
\centering
2870 & 0.958 & 0.812 & 0.576 \\  
957 & 0.924 & 0.833 & 0.646 \\ 
287 & 0.757 & 0.944 & 0.757 \\ 
95.7 & 0.424 & 0.896 & 0.66  \\  
28.7 & 0.056 & 0.799 & 0.486 \\
\midrule
\midrule
\end{tabular}
\label{tab:failureRateNand}
\end{table}

\begin{figure}[h]
	\includegraphics[width=\linewidth]{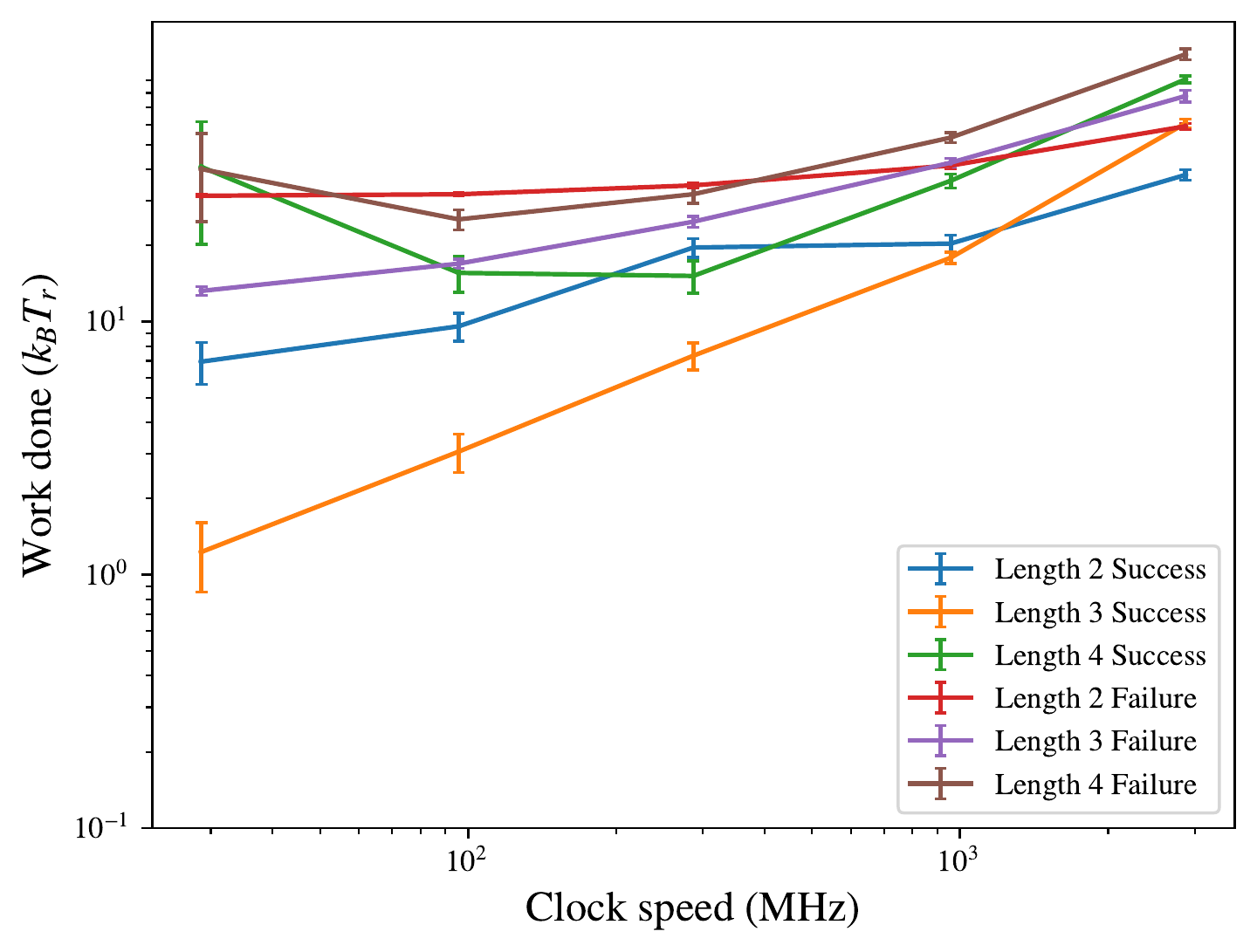}
	\caption{Work done against clock speed for different chain lengths in NAND gates driven via the pipelining protocol for both failed and successful reverse pipelining processes.}
	\label{fig:nandPipeFS}
\end{figure}

\subsection{Pipelined NAND circuits with the modified driver}
\begin{figure*}[h]
	\includegraphics[width=1\linewidth]{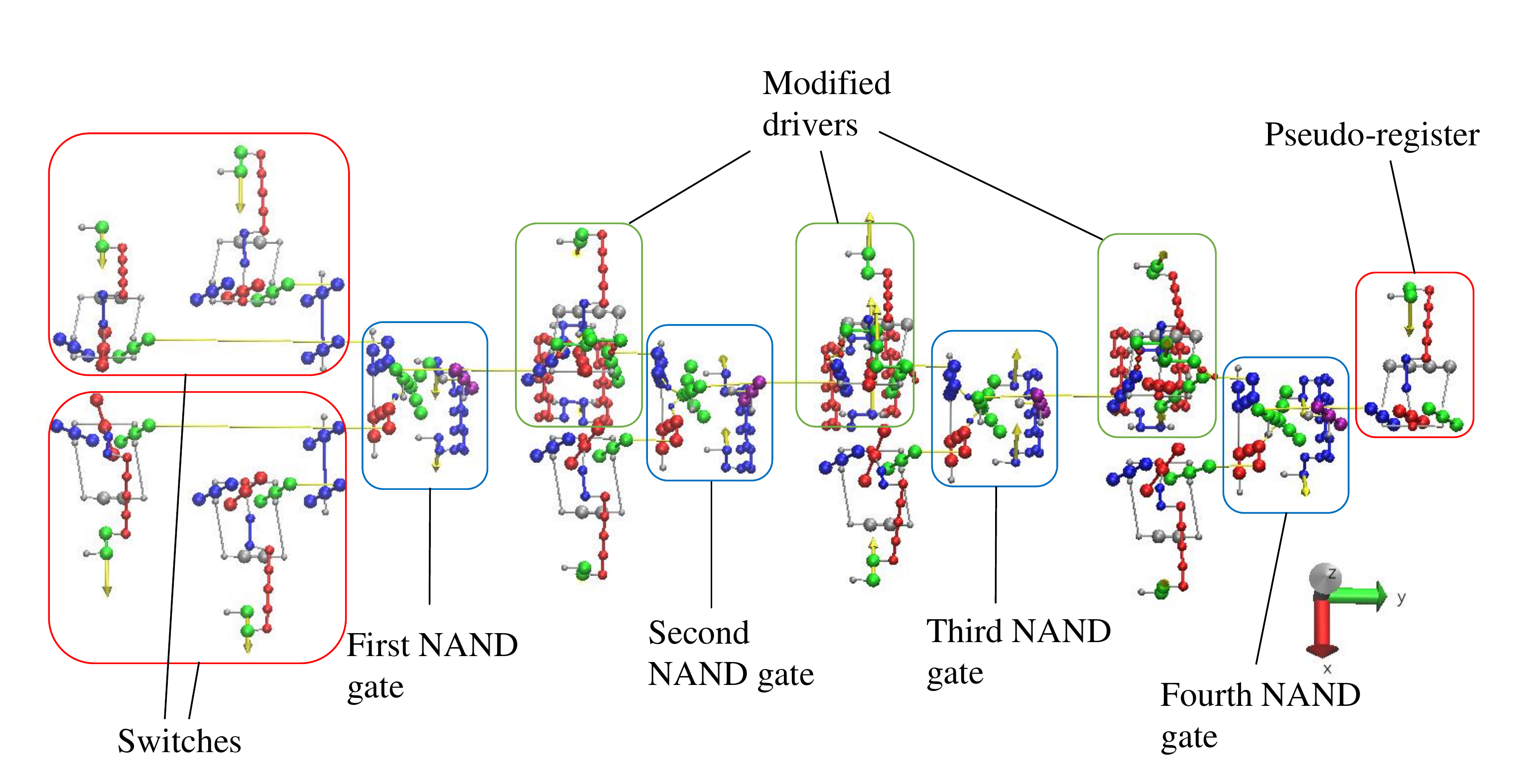}
	\caption{The four-gate pipelined system with all drivers between two NAND gates having being replaced by modified drivers and equipped with a pseudo-register. 
    }
	\label{fig:dipe3Label}
\end{figure*}

Fig.\ \ref{fig:dipe3Label} illustrates the four-gate pipelined system with the modified drivers replacing all drivers that connect two NAND gates.

%% file: apssamp.bbl
\providecommand{\noopsort}[1]{}\providecommand{\singleletter}[1]{#1}%
\begin{thebibliography}{29}%
\makeatletter
\providecommand \@ifxundefined [1]{%
 \@ifx{#1\undefined}
}%
\providecommand \@ifnum [1]{%
 \ifnum #1\expandafter \@firstoftwo
 \else \expandafter \@secondoftwo
 \fi
}%
\providecommand \@ifx [1]{%
 \ifx #1\expandafter \@firstoftwo
 \else \expandafter \@secondoftwo
 \fi
}%
\providecommand \natexlab [1]{#1}%
\providecommand \enquote  [1]{``#1''}%
\providecommand \bibnamefont  [1]{#1}%
\providecommand \bibfnamefont [1]{#1}%
\providecommand \citenamefont [1]{#1}%
\providecommand \href@noop [0]{\@secondoftwo}%
\providecommand \href [0]{\begingroup \@sanitize@url \@href}%
\providecommand \@href[1]{\@@startlink{#1}\@@href}%
\providecommand \@@href[1]{\endgroup#1\@@endlink}%
\providecommand \@sanitize@url [0]{\catcode `\\12\catcode `\$12\catcode
  `\&12\catcode `\#12\catcode `\^12\catcode `\_12\catcode `\%12\relax}%
\providecommand \@@startlink[1]{}%
\providecommand \@@endlink[0]{}%
\providecommand \url  [0]{\begingroup\@sanitize@url \@url }%
\providecommand \@url [1]{\endgroup\@href {#1}{\urlprefix }}%
\providecommand \urlprefix  [0]{URL }%
\providecommand \Eprint [0]{\href }%
\providecommand \doibase [0]{https://doi.org/}%
\providecommand \selectlanguage [0]{\@gobble}%
\providecommand \bibinfo  [0]{\@secondoftwo}%
\providecommand \bibfield  [0]{\@secondoftwo}%
\providecommand \translation [1]{[#1]}%
\providecommand \BibitemOpen [0]{}%
\providecommand \bibitemStop [0]{}%
\providecommand \bibitemNoStop [0]{.\EOS\space}%
\providecommand \EOS [0]{\spacefactor3000\relax}%
\providecommand \BibitemShut  [1]{\csname bibitem#1\endcsname}%
\let\auto@bib@innerbib\@empty
\bibitem [{\citenamefont {Landauer}(1961)}]{Lan61}%
  \BibitemOpen
  \bibfield  {author} {\bibinfo {author} {\bibfnamefont {R.}~\bibnamefont
  {Landauer}},\ }\href@noop {} {\bibfield  {journal} {\bibinfo  {journal} {IBM
  J. Res. Dev.}\ }\textbf {\bibinfo {volume} {5}},\ \bibinfo {pages} {183}
  (\bibinfo {year} {1961})}\BibitemShut {NoStop}%
\bibitem [{\citenamefont {Fredkin}\ and\ \citenamefont
  {Toffoli}(1982)}]{Fred82}%
  \BibitemOpen
  \bibfield  {author} {\bibinfo {author} {\bibfnamefont {R.}~\bibnamefont
  {Fredkin}}\ and\ \bibinfo {author} {\bibfnamefont {T.}~\bibnamefont
  {Toffoli}},\ }\href@noop {} {\bibfield  {journal} {\bibinfo  {journal} {Int.
  J. Theor. Phys.}\ }\textbf {\bibinfo {volume} {21}},\ \bibinfo {pages}
  {219–253} (\bibinfo {year} {1982})}\BibitemShut {NoStop}%
\bibitem [{\citenamefont {Wolpert}(2019)}]{wol19}%
  \BibitemOpen
  \bibfield  {author} {\bibinfo {author} {\bibfnamefont {D.~H.}\ \bibnamefont
  {Wolpert}},\ }\href@noop {} {\bibfield  {journal} {\bibinfo  {journal} {J.
  Phys. A Math. Theor.}\ }\textbf {\bibinfo {volume} {52}},\ \bibinfo {pages}
  {193001} (\bibinfo {year} {2019})}\BibitemShut {NoStop}%
\bibitem [{\citenamefont {Fahn}(1996)}]{fahn96}%
  \BibitemOpen
  \bibfield  {author} {\bibinfo {author} {\bibfnamefont {P.~N.}\ \bibnamefont
  {Fahn}},\ }\href@noop {} {\bibfield  {journal} {\bibinfo  {journal}
  {Foundations of Physics}\ }\textbf {\bibinfo {volume} {26}},\ \bibinfo
  {pages} {71} (\bibinfo {year} {1996})}\BibitemShut {NoStop}%
\bibitem [{\citenamefont {Maroney}(2005)}]{mar05}%
  \BibitemOpen
  \bibfield  {author} {\bibinfo {author} {\bibfnamefont {O.~J.}\ \bibnamefont
  {Maroney}},\ }\href@noop {} {\bibfield  {journal} {\bibinfo  {journal}
  {Studies in History and Philosophy of Science Part B: Studies in History and
  Philosophy of Modern Physics}\ }\textbf {\bibinfo {volume} {36}},\ \bibinfo
  {pages} {355} (\bibinfo {year} {2005})}\BibitemShut {NoStop}%
\bibitem [{\citenamefont {Brittain}\ \emph {et~al.}(2021)\citenamefont
  {Brittain}, \citenamefont {Jones},\ and\ \citenamefont {Ouldridge}}]{Brit21}%
  \BibitemOpen
  \bibfield  {author} {\bibinfo {author} {\bibfnamefont {R.~A.}\ \bibnamefont
  {Brittain}}, \bibinfo {author} {\bibfnamefont {N.~S.}\ \bibnamefont
  {Jones}},\ and\ \bibinfo {author} {\bibfnamefont {T.~E.}\ \bibnamefont
  {Ouldridge}},\ }\href@noop {} {} (\bibinfo {year} {2021}),\ \Eprint
  {https://arxiv.org/abs/2102.03388} {arXiv:2102.03388} \BibitemShut {NoStop}%
\bibitem [{\citenamefont {Ouldridge}\ and\ \citenamefont
  {Wolpert}(2022)}]{ould22}%
  \BibitemOpen
  \bibfield  {author} {\bibinfo {author} {\bibfnamefont {T.~E.}\ \bibnamefont
  {Ouldridge}}\ and\ \bibinfo {author} {\bibfnamefont {D.~H.}\ \bibnamefont
  {Wolpert}},\ }\href@noop {} {\bibfield  {journal} {\bibinfo  {journal} {arXiv
  preprint arXiv:2208.06895}\ } (\bibinfo {year} {2022})}\BibitemShut {NoStop}%
\bibitem [{\citenamefont {Wolpert}\ and\ \citenamefont
  {Kolchinsky}(2020)}]{wol20}%
  \BibitemOpen
  \bibfield  {author} {\bibinfo {author} {\bibfnamefont {D.~H.}\ \bibnamefont
  {Wolpert}}\ and\ \bibinfo {author} {\bibfnamefont {A.}~\bibnamefont
  {Kolchinsky}},\ }\href@noop {} {\bibfield  {journal} {\bibinfo  {journal}
  {New J. Phys.}\ }\textbf {\bibinfo {volume} {22}},\ \bibinfo {pages} {063047}
  (\bibinfo {year} {2020})}\BibitemShut {NoStop}%
\bibitem [{\citenamefont {Seet}\ \emph {et~al.}(2023)\citenamefont {Seet},
  \citenamefont {Ouldridge},\ and\ \citenamefont {Doye}}]{seet22}%
  \BibitemOpen
  \bibfield  {author} {\bibinfo {author} {\bibfnamefont {I.}~\bibnamefont
  {Seet}}, \bibinfo {author} {\bibfnamefont {T.~E.}\ \bibnamefont
  {Ouldridge}},\ and\ \bibinfo {author} {\bibfnamefont {J.~P.~K.}\ \bibnamefont
  {Doye}},\ }\href {https://doi.org/10.1103/PhysRevE.107.024134} {\bibfield
  {journal} {\bibinfo  {journal} {Phys. Rev. E}\ }\textbf {\bibinfo {volume}
  {107}},\ \bibinfo {pages} {024134} (\bibinfo {year} {2023})}\BibitemShut
  {NoStop}%
\bibitem [{\citenamefont {Kolchinsky}\ and\ \citenamefont
  {Wolpert}(2020)}]{Kol20}%
  \BibitemOpen
  \bibfield  {author} {\bibinfo {author} {\bibfnamefont {A.}~\bibnamefont
  {Kolchinsky}}\ and\ \bibinfo {author} {\bibfnamefont {D.~H.}\ \bibnamefont
  {Wolpert}},\ }\href {https://doi.org/10.1103/PhysRevResearch.2.033312}
  {\bibfield  {journal} {\bibinfo  {journal} {Phys. Rev. Research}\ }\textbf
  {\bibinfo {volume} {2}},\ \bibinfo {pages} {033312} (\bibinfo {year}
  {2020})}\BibitemShut {NoStop}%
\bibitem [{\citenamefont {Genot}\ \emph {et~al.}(2011)\citenamefont {Genot},
  \citenamefont {Bath},\ and\ \citenamefont {Tuberfield}}]{Gen11}%
  \BibitemOpen
  \bibfield  {author} {\bibinfo {author} {\bibfnamefont {A.}~\bibnamefont
  {Genot}}, \bibinfo {author} {\bibfnamefont {J.}~\bibnamefont {Bath}},\ and\
  \bibinfo {author} {\bibfnamefont {A.}~\bibnamefont {Tuberfield}},\
  }\href@noop {} {\bibfield  {journal} {\bibinfo  {journal} {J. Am. Chem.
  Soc.}\ }\textbf {\bibinfo {volume} {133}},\ \bibinfo {pages} {20080}
  (\bibinfo {year} {2011})}\BibitemShut {NoStop}%
\bibitem [{\citenamefont {Li}\ \emph {et~al.}(2014)\citenamefont {Li},
  \citenamefont {Lohmann},\ and\ \citenamefont {Famulok}}]{Tao14}%
  \BibitemOpen
  \bibfield  {author} {\bibinfo {author} {\bibfnamefont {T.}~\bibnamefont
  {Li}}, \bibinfo {author} {\bibfnamefont {F.}~\bibnamefont {Lohmann}},\ and\
  \bibinfo {author} {\bibfnamefont {M.}~\bibnamefont {Famulok}},\ }\href@noop
  {} {\bibfield  {journal} {\bibinfo  {journal} {Nat. Commun.}\ }\textbf
  {\bibinfo {volume} {5}},\ \bibinfo {pages} {4940} (\bibinfo {year}
  {2014})}\BibitemShut {NoStop}%
\bibitem [{\citenamefont {Yasuda}\ \emph {et~al.}(2021)\citenamefont {Yasuda},
  \citenamefont {Buskohl}, \citenamefont {Gillman}, \citenamefont {Murphey},
  \citenamefont {Stepney}, \citenamefont {Vaia},\ and\ \citenamefont
  {Raney}}]{Yas21}%
  \BibitemOpen
  \bibfield  {author} {\bibinfo {author} {\bibfnamefont {H.}~\bibnamefont
  {Yasuda}}, \bibinfo {author} {\bibfnamefont {P.~R.}\ \bibnamefont {Buskohl}},
  \bibinfo {author} {\bibfnamefont {A.}~\bibnamefont {Gillman}}, \bibinfo
  {author} {\bibfnamefont {T.~D.}\ \bibnamefont {Murphey}}, \bibinfo {author}
  {\bibfnamefont {S.}~\bibnamefont {Stepney}}, \bibinfo {author} {\bibfnamefont
  {R.~A.}\ \bibnamefont {Vaia}},\ and\ \bibinfo {author} {\bibfnamefont
  {J.~R.}\ \bibnamefont {Raney}},\ }\href@noop {} {\bibfield  {journal}
  {\bibinfo  {journal} {Nature}\ }\textbf {\bibinfo {volume} {598}},\ \bibinfo
  {pages} {39} (\bibinfo {year} {2021})}\BibitemShut {NoStop}%
\bibitem [{\citenamefont {Merkle}(1993)}]{Mer93}%
  \BibitemOpen
  \bibfield  {author} {\bibinfo {author} {\bibfnamefont {R.~C.}\ \bibnamefont
  {Merkle}},\ }\href@noop {} {\bibfield  {journal} {\bibinfo  {journal}
  {Nanotechnology}\ }\textbf {\bibinfo {volume} {4}},\ \bibinfo {pages} {114}
  (\bibinfo {year} {1993})}\BibitemShut {NoStop}%
\bibitem [{\citenamefont {Wenzler}\ \emph {et~al.}(2011)\citenamefont
  {Wenzler}, \citenamefont {Dunn}, \citenamefont {Toffoli},\ and\ \citenamefont
  {Mohanty}}]{Wen11}%
  \BibitemOpen
  \bibfield  {author} {\bibinfo {author} {\bibfnamefont {J.}~\bibnamefont
  {Wenzler}}, \bibinfo {author} {\bibfnamefont {T.}~\bibnamefont {Dunn}},
  \bibinfo {author} {\bibfnamefont {T.}~\bibnamefont {Toffoli}},\ and\ \bibinfo
  {author} {\bibfnamefont {P.}~\bibnamefont {Mohanty}},\ }\href@noop {}
  {\bibfield  {journal} {\bibinfo  {journal} {Nano Lett.}\ }\textbf {\bibinfo
  {volume} {14}},\ \bibinfo {pages} {89} (\bibinfo {year} {2011})}\BibitemShut
  {NoStop}%
\bibitem [{\citenamefont {Merkle}\ \emph {et~al.}(2018)\citenamefont {Merkle},
  \citenamefont {Freitas}, \citenamefont {Hogg}, \citenamefont {Moore},
  \citenamefont {Moses},\ and\ \citenamefont {Ryley}}]{Mer18}%
  \BibitemOpen
  \bibfield  {author} {\bibinfo {author} {\bibfnamefont {R.~C.}\ \bibnamefont
  {Merkle}}, \bibinfo {author} {\bibfnamefont {R.~A.}\ \bibnamefont {Freitas}},
  \bibinfo {author} {\bibfnamefont {T.}~\bibnamefont {Hogg}}, \bibinfo {author}
  {\bibfnamefont {T.~E.}\ \bibnamefont {Moore}}, \bibinfo {author}
  {\bibfnamefont {M.~S.}\ \bibnamefont {Moses}},\ and\ \bibinfo {author}
  {\bibfnamefont {J.}~\bibnamefont {Ryley}},\ }\href@noop {} {\bibfield
  {journal} {\bibinfo  {journal} {J. Mechanisms Robotics.}\ }\textbf {\bibinfo
  {volume} {10}},\ \bibinfo {pages} {061006} (\bibinfo {year}
  {2018})}\BibitemShut {NoStop}%
\bibitem [{\citenamefont {Song}\ \emph {et~al.}(2019)\citenamefont {Song},
  \citenamefont {Panas}, \citenamefont {Chizari}, \citenamefont {Shaw},
  \citenamefont {Jackson}, \citenamefont {Hopkins},\ and\ \citenamefont
  {Pascall}}]{Song19}%
  \BibitemOpen
  \bibfield  {author} {\bibinfo {author} {\bibfnamefont {Y.}~\bibnamefont
  {Song}}, \bibinfo {author} {\bibfnamefont {R.~M.}\ \bibnamefont {Panas}},
  \bibinfo {author} {\bibfnamefont {S.}~\bibnamefont {Chizari}}, \bibinfo
  {author} {\bibfnamefont {L.~A.}\ \bibnamefont {Shaw}}, \bibinfo {author}
  {\bibfnamefont {J.~A.}\ \bibnamefont {Jackson}}, \bibinfo {author}
  {\bibfnamefont {J.~B.}\ \bibnamefont {Hopkins}},\ and\ \bibinfo {author}
  {\bibfnamefont {A.~J.}\ \bibnamefont {Pascall}},\ }\href@noop {} {\bibfield
  {journal} {\bibinfo  {journal} {Nat. Commun.}\ }\textbf {\bibinfo {volume}
  {10}},\ \bibinfo {pages} {882} (\bibinfo {year} {2019})}\BibitemShut
  {NoStop}%
\bibitem [{\citenamefont {Be\v{c}v\'{a}\v{r}}(2004)}]{bev04}%
  \BibitemOpen
  \bibfield  {author} {\bibinfo {author} {\bibfnamefont {M.}~\bibnamefont
  {Be\v{c}v\'{a}\v{r}}}\ }(\bibinfo  {publisher} {Association for Computing
  Machinery},\ \bibinfo {address} {New York, NY, USA},\ \bibinfo {year}
  {2004})\ p.\ \bibinfo {pages} {16–es}\BibitemShut {NoStop}%
\bibitem [{\citenamefont {Wong}\ \emph {et~al.}(1993)\citenamefont {Wong},
  \citenamefont {De~Micheli},\ and\ \citenamefont {Flynn}}]{Wong93}%
  \BibitemOpen
  \bibfield  {author} {\bibinfo {author} {\bibfnamefont {D.}~\bibnamefont
  {Wong}}, \bibinfo {author} {\bibfnamefont {G.}~\bibnamefont {De~Micheli}},\
  and\ \bibinfo {author} {\bibfnamefont {M.}~\bibnamefont {Flynn}},\ }\href
  {https://doi.org/10.1109/43.184841} {\bibfield  {journal} {\bibinfo
  {journal} {IEEE Trans. Comput.-Aided Des. Integr. Circuits Syst.}\ }\textbf
  {\bibinfo {volume} {12}},\ \bibinfo {pages} {25} (\bibinfo {year}
  {1993})}\BibitemShut {NoStop}%
\bibitem [{\citenamefont {Burleson}\ \emph {et~al.}(1998)\citenamefont
  {Burleson}, \citenamefont {Ciesielski}, \citenamefont {Klass},\ and\
  \citenamefont {Liu}}]{Bur98}%
  \BibitemOpen
  \bibfield  {author} {\bibinfo {author} {\bibfnamefont {W.~P.}\ \bibnamefont
  {Burleson}}, \bibinfo {author} {\bibfnamefont {M.}~\bibnamefont
  {Ciesielski}}, \bibinfo {author} {\bibfnamefont {F.}~\bibnamefont {Klass}},\
  and\ \bibinfo {author} {\bibfnamefont {W.}~\bibnamefont {Liu}},\ }\href@noop
  {} {\bibfield  {journal} {\bibinfo  {journal} {IEEE Trans. Very Large Scale
  Integr. (VLSI) Syst.}\ }\textbf {\bibinfo {volume} {6}},\ \bibinfo {pages}
  {464} (\bibinfo {year} {1998})}\BibitemShut {NoStop}%
\bibitem [{\citenamefont {Gray}\ \emph {et~al.}(1994)\citenamefont {Gray},
  \citenamefont {Liu}, \citenamefont {Cavin~III},\ and\ \citenamefont
  {Cavin}}]{gray94}%
  \BibitemOpen
  \bibfield  {author} {\bibinfo {author} {\bibfnamefont {C.~T.}\ \bibnamefont
  {Gray}}, \bibinfo {author} {\bibfnamefont {W.}~\bibnamefont {Liu}}, \bibinfo
  {author} {\bibfnamefont {R.~K.}\ \bibnamefont {Cavin~III}},\ and\ \bibinfo
  {author} {\bibfnamefont {R.~K.}\ \bibnamefont {Cavin}},\ }\href@noop {}
  {\emph {\bibinfo {title} {Wave Pipelining: Theory and CMOS
  Implementation}}},\ Vol.\ \bibinfo {volume} {248}\ (\bibinfo  {publisher}
  {Springer Science \& Business Media},\ \bibinfo {year} {1994})\BibitemShut
  {NoStop}%
\bibitem [{\citenamefont {Shen}\ and\ \citenamefont {Lipasti}(2005)}]{shen05}%
  \BibitemOpen
  \bibfield  {author} {\bibinfo {author} {\bibfnamefont {J.}~\bibnamefont
  {Shen}}\ and\ \bibinfo {author} {\bibfnamefont {M.}~\bibnamefont {Lipasti}},\
  }\href {https://books.google.co.uk/books?id=Nibfj2aXwLYC} {\emph {\bibinfo
  {title} {Modern Processor Design: Fundamentals of Superscalar Processors}}},\
  Electrical and Computer Engineering\ (\bibinfo  {publisher} {McGraw-Hill
  Companies, Incorporated},\ \bibinfo {year} {2005})\BibitemShut {NoStop}%
\bibitem [{\citenamefont {Davidchack}\ \emph {et~al.}(2015)\citenamefont
  {Davidchack}, \citenamefont {Ouldridge},\ and\ \citenamefont
  {Tretyakov}}]{Dav14}%
  \BibitemOpen
  \bibfield  {author} {\bibinfo {author} {\bibfnamefont {R.~L.}\ \bibnamefont
  {Davidchack}}, \bibinfo {author} {\bibfnamefont {T.~E.}\ \bibnamefont
  {Ouldridge}},\ and\ \bibinfo {author} {\bibfnamefont {M.~V.}\ \bibnamefont
  {Tretyakov}},\ }\href@noop {} {\bibfield  {journal} {\bibinfo  {journal} {J.
  Chem. Phys.}\ }\textbf {\bibinfo {volume} {142}},\ \bibinfo {pages} {144114}
  (\bibinfo {year} {2015})}\BibitemShut {NoStop}%
\bibitem [{\citenamefont {Sekimoto}(1997)}]{Sek97}%
  \BibitemOpen
  \bibfield  {author} {\bibinfo {author} {\bibfnamefont {K.}~\bibnamefont
  {Sekimoto}},\ }\href {https://doi.org/10.1143/JPSJ.66.1234} {\bibfield
  {journal} {\bibinfo  {journal} {Journal of the Physical Society of Japan}\
  }\textbf {\bibinfo {volume} {66}},\ \bibinfo {pages} {1234} (\bibinfo {year}
  {1997})}\BibitemShut {NoStop}%
\bibitem [{\citenamefont {Sekimoto}(1998)}]{Sek98}%
  \BibitemOpen
  \bibfield  {author} {\bibinfo {author} {\bibfnamefont {K.}~\bibnamefont
  {Sekimoto}},\ }\href {https://doi.org/10.1143/PTPS.130.17} {\bibfield
  {journal} {\bibinfo  {journal} {Prog. Theor. Phys.}\ }\textbf {\bibinfo
  {volume} {130}},\ \bibinfo {pages} {17} (\bibinfo {year} {1998})}\BibitemShut
  {NoStop}%
\bibitem [{\citenamefont {Jun}\ \emph {et~al.}(2014)\citenamefont {Jun},
  \citenamefont {Gavrilov},\ and\ \citenamefont {Bechhoefer}}]{jun14}%
  \BibitemOpen
  \bibfield  {author} {\bibinfo {author} {\bibfnamefont {Y.}~\bibnamefont
  {Jun}}, \bibinfo {author} {\bibfnamefont {M.}~\bibnamefont {Gavrilov}},\ and\
  \bibinfo {author} {\bibfnamefont {J.}~\bibnamefont {Bechhoefer}},\
  }\href@noop {} {\bibfield  {journal} {\bibinfo  {journal} {Physical review
  letters}\ }\textbf {\bibinfo {volume} {113}},\ \bibinfo {pages} {190601}
  (\bibinfo {year} {2014})}\BibitemShut {NoStop}%
\bibitem [{\citenamefont {Ouldridge}\ \emph {et~al.}(2017)\citenamefont
  {Ouldridge}, \citenamefont {Govern},\ and\ \citenamefont {ten
  Wolde}}]{ould17}%
  \BibitemOpen
  \bibfield  {author} {\bibinfo {author} {\bibfnamefont {T.~E.}\ \bibnamefont
  {Ouldridge}}, \bibinfo {author} {\bibfnamefont {C.~C.}\ \bibnamefont
  {Govern}},\ and\ \bibinfo {author} {\bibfnamefont {P.~R.}\ \bibnamefont {ten
  Wolde}},\ }\href@noop {} {\bibfield  {journal} {\bibinfo  {journal} {Physical
  Review X}\ }\textbf {\bibinfo {volume} {7}},\ \bibinfo {pages} {021004}
  (\bibinfo {year} {2017})}\BibitemShut {NoStop}%
\bibitem [{\citenamefont {Riechers}\ \emph {et~al.}(2020)\citenamefont
  {Riechers}, \citenamefont {Boyd}, \citenamefont {Wimsatt},\ and\
  \citenamefont {Crutchfield}}]{rie20}%
  \BibitemOpen
  \bibfield  {author} {\bibinfo {author} {\bibfnamefont {P.~M.}\ \bibnamefont
  {Riechers}}, \bibinfo {author} {\bibfnamefont {A.~B.}\ \bibnamefont {Boyd}},
  \bibinfo {author} {\bibfnamefont {G.~W.}\ \bibnamefont {Wimsatt}},\ and\
  \bibinfo {author} {\bibfnamefont {J.~P.}\ \bibnamefont {Crutchfield}},\
  }\href@noop {} {\bibfield  {journal} {\bibinfo  {journal} {Physical Review
  Research}\ }\textbf {\bibinfo {volume} {2}},\ \bibinfo {pages} {033524}
  (\bibinfo {year} {2020})}\BibitemShut {NoStop}%
\bibitem [{\citenamefont {Esposito}\ and\ \citenamefont {Van~den
  Broeck}(2011)}]{esp11}%
  \BibitemOpen
  \bibfield  {author} {\bibinfo {author} {\bibfnamefont {M.}~\bibnamefont
  {Esposito}}\ and\ \bibinfo {author} {\bibfnamefont {C.}~\bibnamefont {Van~den
  Broeck}},\ }\href@noop {} {\bibfield  {journal} {\bibinfo  {journal}
  {Europhysics Letters}\ }\textbf {\bibinfo {volume} {95}},\ \bibinfo {pages}
  {40004} (\bibinfo {year} {2011})}\BibitemShut {NoStop}%
\end{thebibliography}%
